\documentclass[reprint,amsmath,amssymb,aps,twocolumn]{revtex4-2}

\usepackage{graphicx}
\usepackage{bm, slashed}
\usepackage{tikz} 
\usepackage{simpler-wick}
\usepackage{hyperref} 
\usepackage{comment}
\usepackage{mathdots,bbold}
\usepackage{soul}
\usepackage[normalem]{ulem}
\setcitestyle{square,numbers}
\usepackage{pgfplotstable}

\begin{document}

    \title{Superconducting properties of the three-dimensional Hofstadter--Hubbard model below the critical flux for Weyl points}
    
    \author{Pierpaolo Fontana}
    \email{Pierpaolo.Fontana@uab.cat}
    \affiliation{
        Departament de Física, Universitat Aut\`{o}noma de Barcelona, 08193 Bellaterra, Spain.
    }
    
    \author{Luca Lepori}
    \affiliation{Dipartimento di Scienze Matematiche, Fisiche e Informatiche, Universit\`a  di Parma, Parco Area delle Scienze, 53/A, I-43124 Parma, Italy.}
    \affiliation{Gruppo Collegato di Parma, INFN-Sezione Milano-Bicocca, I-43124 Parma, Italy}
    \affiliation{UdR Parma, INSTM, I-43124 Parma, Italy.}

    \author{Andrea Trombettoni}
    \affiliation{%
        Department of Physics, University of Trieste, Strada Costiera 11, I-34151 Trieste, Italy
    }
    \affiliation{SISSA and INFN, Sezione di Trieste, Via Bonomea 265,
        I-34136 Trieste, Italy}
    
    \date{\today}
    
    \begin{abstract}
        The three-dimensional Hofstadter model exhibits a critical rational flux at which Weyl points emerge in the single-particle spectrum. We study the superconducting regime of the model in the presence of a Hubbard attractive interaction by tuning the magnetic flux across its critical value. We determine the phase diagram in the plane of the coprime pairs parametrizing the magnetic flux. We show that the system exhibits two distinct regimes separated by a critical flux $\Phi_c$: for $\Phi>\Phi_c$, a semimetal-to-superconductor quantum phase transition occurs at a finite interaction strength ($U_c\neq0$), while for $\Phi<\Phi_c$ superconductivity arises for arbitrarily weak attraction, with a BCS-like exponential scaling of the gap due to the finiteness of the density of states. Close to the transition, we study the scaling behavior and identify the critical exponents. Our results highlight the interplay between magnetic band topology and attractive pairing in three-dimensional Hofstadter systems.
    \end{abstract}
    
    \maketitle
    
    \section{\label{intro}Introduction}

    The realization of artificial gauge potentials in ultracold atomic gases has significantly expanded the aim of quantum simulation. By engineering laser-induced couplings between internal atomic states, it is possible to generate synthetic Abelian and non-Abelian gauge fields acting on neutral atoms \cite{lewenstein2012,fallanibook,goldman2014,yagomalo2024,Aidelsburger2016}. These techniques have enabled the realization of effective electromagnetic fields \cite{Aidelsburger2018}, spin-orbit coupling \cite{Galitski2013,Huang2016}, and complex tunneling amplitudes in fermionic and bosonic systems confined in optical lattices \cite{Eckardt2017,Weitenberg2021}. In lattice systems, artificial gauge potentials manifest as Peierls phases \cite{Peierls1933} acquired in tunneling processes, producing complex hopping amplitudes and accumulating uniform magnetic fluxes when a particle closes a lattice plaquette. Such control has established ultracold atoms as a versatile platform for exploring topological band structures and strongly correlated quantum matter \cite{CooperRMP2019,yagomalo2024}. Similar phenomena have also been investigated in other engineered quantum platforms, including photonic crystals \cite{ozawa2019}, Moiré heterostructures \cite{Dean2013, Hunt2013}, and molecular nanostructures \cite{kempkes2019,fremling2020}, highlighting the broad relevance of synthetic gauge fields beyond the realm of ultracold atomic systems.

    A central model in these contexts is the Harper--Hofstadter model, often simply referred to as the Hofstadter model, which describes particles hopping on a lattice in the presence of a uniform magnetic flux \cite{Hubbard1963,HofstadterPRB1976}. Experimentally, low-dimensional versions of the model have been realized using Raman-assisted tunneling and Floquet engineering in optical lattices \cite{Aidelsburger2013,Miyake2013,Kennedy2013,Kennedy2015,Aidelsburger2015,Stuhl2015,Tai2017}, enabling direct access to magnetic Bloch bands and fractal energy spectra \cite{fallaniribbon}. In two dimensions, the Hofstadter model exhibits the celebrated butterfly structure \cite{HofstadterPRB1976} and supports Chern bands associated with the integer quantum Hall effect \cite{ludwig2009}. A time-reversal-preserving model has also been proposed \cite{IskinPRA2016,IskinPRL2017}, providing lattice analogs of quantum spin-Hall systems. For particular flux values, Dirac cones emerge in the magnetic Brillouin zone \cite{Affleck-MarstonPRB1989,Mazzucchi_2013}, linking Hofstadter physics to graphene-like features \cite{El-Batanouny2020}.

    The inclusion of Hubbard interactions significantly enriches the scenario provided by the Hofstadter Hamiltonian, leading to the so-called Hofstadter–Hubbard (HH) model. In the repulsive regime of the Hubbard interaction, competition arises between insulating or magnetic phases driven by correlations and the underlying magnetic band structure \cite{JackschPRL1998,Greiner2002,UmucalilarPRA2007,HafeziPRA2007,GoldbaumPRA2008,UmucalilarPRA2010,Shaffer2022,SahayPRB2024}. On the other hand, in the attractive regime, pairing instabilities lead to superconducting (SC) states, whose structure reflects the Hofstadter magnetic bands \cite{ZhaiPRL2010,IskinPRA2015,IskinPRA2016, ZengPRL2019}. We remark that, although in this paper we focus on SC properties of the attractive HH model, our results for neutral ultracold fermions also describe their {\it superfluid} properties in presence of attractive interactions between fermions in a two-component ultracold Fermi gas.
    
    In two-dimensional systems, the attractive HH model has been shown to host inhomogeneous SC phases, vortex-like order parameters, and transitions from topological phases to paired states \cite{WangPRB2014,Peotta2015,AnzaiPRA2017}. In this dimension, the interplay between topological effects and superconductivity, referred to as Hofstadter superconductivity, has been extensively investigated, see \cite{SantosPRB2021} and references therein. 

    Despite this progress in two dimensions, the three-dimensional HH model remains comparatively unexplored. The extension to three dimensions introduces additional complexities related to the single-particle band structure \cite{KunsztZeePRB1991,KoshinoPRL2001,KoshinoPRB2003}, modifying the properties of the density of states (DOS) in a flux-dependent way \cite{LaughlinZouPRB1990,Hasegawa1990,Hasegawa1992, BurrelloJPHYSMATH2017,FontanaPRB2021,FontanaPRB2024}. In particular, for a cubic lattice with an isotropic magnetic field oriented along the body diagonal, and flux parametrized as $\Phi=2\pi m/n$, with $m$ and $n$ coprime, the single-particle spectrum exhibits a rich structure that depends sensitively on the rational flux. In the recent work of Ref. \cite{FontanaPRB2024} it was shown that, at fixed $m$, there exists a critical flux $\Phi_c(m)$ that separates two distinct regimes. For $\Phi<\Phi_c(m)$, the magnetic bands completely overlap, resulting in a DOS without isolated zeros. For $\Phi>\Phi_c(m)$, instead, the spectrum develops isolated band-touching points and Weyl nodes between the $m$-th and $(m+1)$-th bands, signaling the emergence of three-dimensional semimetallic features. The same work also provided evidence for a finite asymptotic critical flux $\Phi_c$ in the limit $m\rightarrow\infty$, obtained its value numerically ($\Phi_c/2\pi \approx 0.1296$) and proposed two analytic conjectures, obtaining $\Phi_c^{\rm{(conj)}}/2\pi=7/54$. A natural question then is the characterization of the superconducting or superfluid properties near the critical flux $\Phi_c(m)$.

    \begin{figure}[t]
		\centering
		\includegraphics[width=1.0\linewidth]{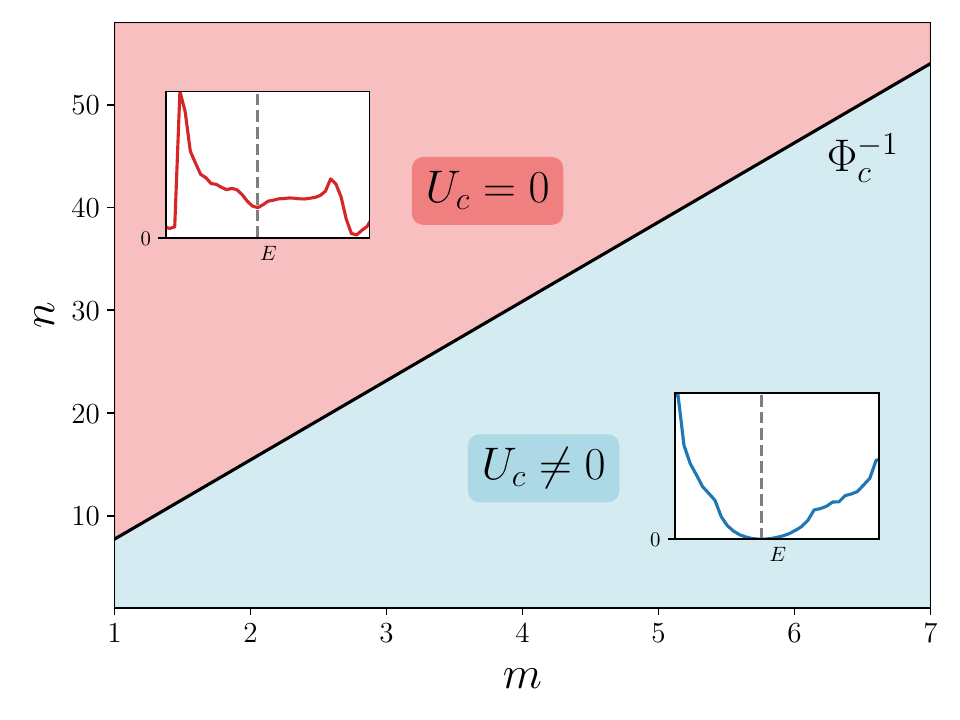}
		\caption{Phase diagram in the plane of the integer coprime pairs $(m,n)$. For $m/n>\Phi_c/2\pi$ (blue region) we identify a critical interaction strength $U_c\neq0$, while in the opposite regime (red region) $U_c=0$. The critical line of $n_c/m_c=2\pi/\phi_c$ (black line) separates the two regimes. In the insets we show an example of the DOS in the two regimes, highlighting their values at the rational fillings $\nu=m/n$ (dashed gray lines). The example values are respectively $\nu=1/3$ (blue region) and $\nu=1/9$ (red region).}
		\label{n_vs_m_phasediagram_Uc}
	\end{figure}
    
    The results for the three-dimensional non-interacting Hofstadter model imply a flux-controlled transition between topologically trivial and nontrivial regimes \cite{Louvet2018,FontanaPRB2021}. Consequently, the pairing problem must take into account the behavior of the DOS near the Fermi level, the location of band extrema, and the presence of Weyl nodes, since these features influence the structure and symmetry of possible SC instabilities. These aspects are directly relevant for SC phenomena in high-field solid-state systems, such as the field-boosted SC observed in UTe$_2$ \cite{Lewin_2023}. The resulting analysis is substantially more intricate than in conventional three-dimensional Hubbard models and is directly shaped by the band topology of the underlying Hofstadter spectrum \cite{SantosPRB2021}.
    
    In this work, we present a study of the attractive three-dimensional HH model. We show how the coupled self-consistent mean-field (MF) equations for the gap and the particle number are modified in the presence of the magnetic flux, and we solve them numerically both in momentum and energy spaces. This approach enables us to map out the SC phase structure in the coprime pairs plane $(m,n)$, as depicted in Fig. \ref{n_vs_m_phasediagram_Uc}, highlighting the role of the critical flux line $\Phi=\Phi_c(m\rightarrow\infty)$ as a separation of topologically trivial and nontrivial regimes even in the presence of Hubbard interactions. We focus on  rational fillings $\nu=m/n$, at which the DOS vanishes at the Fermi level for $n>n_c$, the integer that, at fixed $m$, characterizes the critical flux $\Phi_c(m)$ \cite{FontanaPRB2024}\footnote{At fillings different from $m/n$, the DOS is always finite and the SC instability occurs for arbitrarily small attractive interaction, leading to a trivial $U_c$.}. For fluxes $\Phi>\Phi_c$, we identify a semimetal-superconductor quantum phase transition at finite interaction strength ($U_c\neq0$). In the opposite regime ($\Phi<\Phi_c$), any finite attractive interaction is enough to stabilize a gap, which vanishes exponentially with the inverse of the coupling strength, as for the standard BCS case. The two regimes are separated by the critical flux line ($\Phi=\Phi_c$). We complement the numerical results with an analytical scaling analysis of the pairing field and of the chemical potential close to the transition point.

    The paper is organized as follows. In Sec. \ref{HH_model_section} we introduce the HH model, summarizing the features of the single-particle problem and recalling the MF approximation to derive the self-consistent equations for the gap and the particle number. In Sec. \ref{results_section} we present the numerical results for the pairing field and for the chemical potential, exploring the various regions of the phase diagram, reported in Fig.\ref{n_vs_m_phasediagram_Uc}. In Sec. \ref{scaling_Uc_phi_subsection} we focus on the scaling properties close to the quantum phase transition between semimetal and superconductor, investigating the critical exponents and the interaction strength as functions of the magnetic flux. In Sec. \ref{conclusions} we summarize our conclusions and outlooks. In the Appendices we include technical details about the diagonalization of the single-particle problem (Appendix \ref{Hofstadter_model_appendix} and \ref{Hasegawa_implicit_relations}), the numerical methods employed to solve the self-consistent equations (Appendix \ref{Newton_method_appendix}), a discussion of the results with coprime pairs with $m>1$ (Appendix \ref{mneq1_results}), and the analytical derivation of the scaling relations (Appendices  \ref{pairing_field_scaling} and \ref{shifted_mu_scaling}).
	
	\section{\label{HH_model_section}The Hofstadter--Hubbard model}

\begin{figure}[t!]
		\centering
		\includegraphics[width=1.0\linewidth]{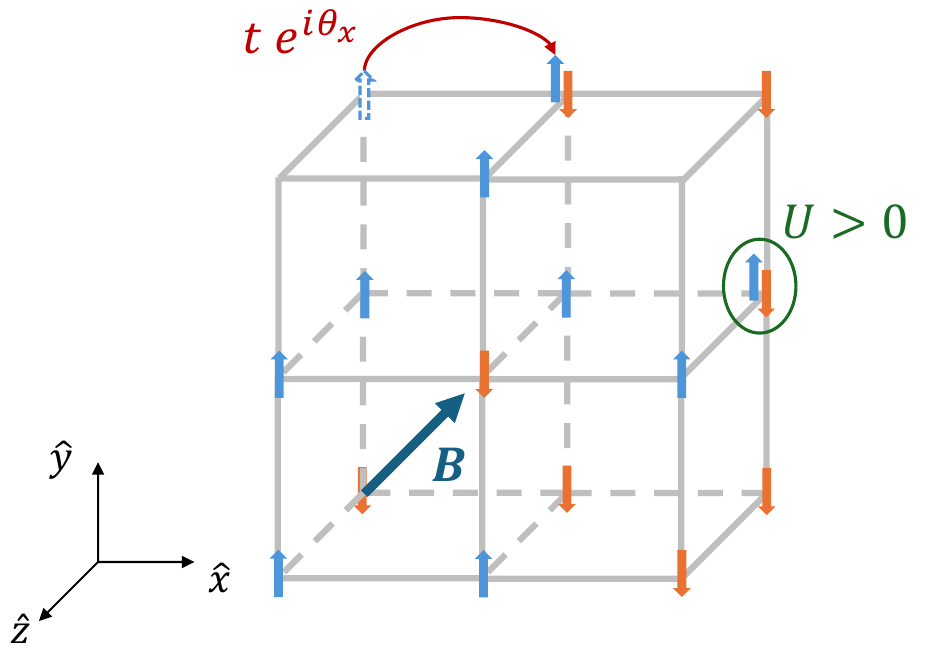}
		\caption{Example of lattice set-up for the three-dimensional Hofstadter--Hubbard model. Fermions are represented as pseudo-spins (blue and orange arrows for up and down spins, respectively). The magnetic field ${\bf B}\parallel\left(1,1,1\right)$. We depict the tunneling along $\hat{x}$ (red arrow) and the on-site Hubbard interaction in the picture (green circle).}
		\label{3d_lattice_def}
	\end{figure}

	We consider the three-dimensional attractive Hofstadter--Hubbard (HH) model on a cubic lattice of $V=L^3$ sites and lattice spacing $a$, with periodic boundary conditions (PBC). The model Hamiltonian in the grand canonical ensemble is
	\begin{align}
		\nonumber
		\mathcal{H}-\mu\mathcal{N}=&-t\sum_{\mathbf{r},\hat{j},\sigma}(c^\dag_{\mathbf{r}+\hat{j},\sigma}e^{i\theta_{\hat{j}}(\mathbf{r})}c_{\mathbf{r}\sigma}+\mathrm{H.c.}) - \\
		&-U\sum_{\mathbf{r}}c^\dag_{\mathbf{r}\uparrow}c^\dag_{\mathbf{r}\downarrow}c_{\mathbf{r}\downarrow}c_{\mathbf{r}\uparrow}-\mu\sum_{\mathbf{r}\sigma}c^\dag_{\mathbf{r}\sigma}c_{\mathbf{r}\sigma} \, .
		\label{3d_hubbard_hofst_hamiltonian}
	\end{align}
	Here, $c_{{\bf r}\sigma}$ and $c^\dagger_{{\bf r}\sigma}$ are the annihilation and creation operators for fermionic particles at lattice site ${\bf r}$ with  pseudo-spin $\sigma=\uparrow,\downarrow$, while $\mu$ is the chemical potential. The strength of the interaction is chosen to be positive, that is, $U>0$, corresponding to on-site attractions. We show an example of the lattice configuration in Fig. \ref{3d_lattice_def}.
    
	\subsection{\label{hofstadter_properties}Properties of the single particle Hamiltonian}
	The free part, when $U=0$, namely
	\begin{equation}
		\mathcal{H}_{\text{Hofst}}=-t\sum_{\mathbf{r},\hat{j},\sigma}c^\dag_{\mathbf{r}+\hat{j},\sigma}e^{i\theta_{\hat{j}}(\mathbf{r})}c_{\mathbf{r}\sigma}+\mathrm{H.c.}
		\label{H_Hofst_def}
	\end{equation}
	is the Hofstadter Hamiltonian \cite{HarperPPS1955,HofstadterPRB1976}. It describes single particles hopping in the cubic lattice in the presence of a static gauge field, whose effects are encoded into the Peierls phase
	\begin{equation}
		\theta_j({\bf r})=\int_{\bf r}^{{\bf r}+\hat{j}} {\bf A}({\bf x})\cdot d{\bf x} \, ,
	\end{equation}
	where $\hat{j}=\{\hat{x},\hat{y},\hat{z}\}$, defining the flux piercing all plaquettes of the cube \cite{Grosso_Parravicini_2000,Peierls1933}. We consider commensurate magnetic fluxes, parametrized by 
    \begin{equation}
        \Phi=2\pi \frac{m}{n},
        \label{phi_def}
        \end{equation}
    with $(m,n)$ coprime integers, and focus on the isotropic case, with the magnetic field oriented along the diagonal of the lattice, i.e., ${\bf B}=\Phi(1,1,1)$, where $\Phi=B a^2$ is the relation between the magnitude of the magnetic field and the commensurate flux.
	
	With these assumptions, the model can be treated in momentum space by introducing the magnetic translation group, as a consequence of the interplay between gauge and translational symmetries in the presence of commensurate background magnetic fields \cite{ZakIPRA1964,ZakIIPRA1964}. A convenient gauge choice for the three-dimensional isotropic problem is given by the Hasegawa gauge \cite{Hasegawa1990} whose vector potential is defined as 	
    \begin{equation}
        {\bf A}({\bf x})=\Phi(0,x-y,y-x) \, .
        \label{gauge}
    \end{equation}
     In reciprocal space, the momentum ${\bf k}$ is restricted to the magnetic Brillouin zone (MBZ), defined as $k_{x,y}\in[-\pi/n,\pi/n]$, $k_z\in[-\pi,\pi]$. This gauge choice allows the reduction of the Hamiltonian problem to the diagonalization of an $n\times n$ matrix for each ${\bf k}\in \text{MBZ}$. The eigenvalues $\epsilon_{\alpha}({\bf k}),\;\alpha=0,\ldots, n-1$ are called magnetic bands, each one being $n$-fold degenerate. We refer to Appendix \ref{Hofstadter_model_appendix} for a summary of these properties, and finally note that this set-up allows for an efficient and structured numerical exact diagonalization (ED) of the Hamiltonian \eqref{H_Hofst_def}. As discussed in \cite{BurrelloJPHYSMATH2017}, the Hasegawa gauge provides an optimal choice, since it reduces the problem to the diagonalization of an $n\times n$ matrix.
    	
	The general properties of the three-dimensional isotropic Hofstadter model have been recently investigated  \cite{FontanaPRB2024}. Through reduced momentum space ED, a critical flux $\Phi_c(m)=2\pi m/n(m)$ has been identified, signaling a transition between two regimes: for $\Phi<\Phi_c$, the energy bands overlap, while in the opposite limit, for $\Phi>\Phi_c$, isolated Weyl points appear between the $m$-th and $(m+1)$-th bands. The critical flux converges to a finite value as $m\rightarrow \infty$, with asymptotic estimate $\Phi_c/2\pi\approx 0.1296$ and a conjectured rational limit of $\Phi^{(\text{conj})}_c/2\pi=7/54$.
	
	\subsection{\label{MF theory_subsection}Hartree--Fock--BCS  approximation}
    
	When $U\neq0$, we treat the attractive HH model in the MF approximation \cite{El-Batanouny2020, Grosso_Parravicini_2000,Altland_Simons_2010}, and consider the real space pairing Hamiltonian
	\begin{align}
		\nonumber
		\mathcal{H}=&\mathcal{H}_{\text{Hofst}}-U\sum_{\mathbf{r},\sigma}\langle 	c^\dag_{\mathbf{r}\sigma}c_{\mathbf{r}\sigma}\rangle c^\dag_{\mathbf{r}-\sigma}c_{\mathbf{r}-\sigma}-\mu\sum_{\mathbf{r}\sigma}c^\dag_{\mathbf{r}\sigma}c_{\mathbf{r}\sigma}\\
		&-\sum_{\mathbf{r}}(\Delta_{{\bf r}}c^\dag_{\mathbf{r}\uparrow}c^\dag_{\mathbf{r}\downarrow}+\Delta^*_{{\bf r}}c_{\mathbf{r}\downarrow}c_{\mathbf{r}\uparrow}) \, .
		\label{real_space_H_pairing}
	\end{align}
	In this expression we retain the terms preserving translational and spin rotational symmetries, implying that $\langle c^\dag_{\mathbf{r}\sigma}c_{\mathbf{r}\sigma'} \rangle=0$ for $\sigma\neq\sigma'$. We consider the presence of a non-trivial SC order parameter, defining 
	\begin{equation}
		\Delta^*_{{\bf r}}\equiv U\langle c^\dag_{\mathbf{r}\uparrow}c^\dag_{\mathbf{r}\downarrow} \rangle \, ,
		\label{gap_definition}
	\end{equation}
	which is non-zero when the global $U(1)$ symmetry is broken in the SC state of the system. Finally, due to the spin rotational symmetry we have that
	\begin{equation}
		\langle c^\dag_{\mathbf{r}\uparrow}c_{\mathbf{r}\uparrow}\rangle=\langle c^\dag_{\mathbf{r}\downarrow}c_{\mathbf{r}\downarrow}\rangle=\frac{f}{2} \, ,
	\end{equation}
	with $f=N/V$ being the lattice filling, i.e. the number of particles per sites in the system.
	
	The pairing Hamiltonian can be written in momentum space in a completely general way, taking into account the proper structure of the magnetic translation group and considering all the possible pairing channels \cite{IskinPRA2016,SantosPRB2021}. Among the possible scenarios, here we consider an homogeneous SC phase, whose order parameter is uniform in orbital space and can be described through a single pairing field $\Delta$ \cite{IskinPRL2017}. This assumption is motivated by the rotational invariance of the attractive Hubbard interaction. We introduce the Nambu spinors 
	\begin{equation}
		\Psi_{{\bf k}}=
		\begin{pmatrix}
			C_{{\bf k}\uparrow}\\
			C_{-{\bf k}\downarrow}
		\end{pmatrix},
		\qquad
		\Psi^\dagger_{{\bf k}}=
		\begin{pmatrix}
			C^\dagger_{{\bf k}\uparrow} & C^\dagger_{-{\bf k}\downarrow} \, ,
		\end{pmatrix}
	\end{equation}
	where $C^\dagger_{{\bf k}}=(c^\dagger_{0{\bf k}},\ldots,c^\dagger_{{n-1},{\bf k}})$ is the vector of operators $c^\dagger_{\alpha{\bf k}}$, with ${\bf k}\in\text{MBZ}$ and patch index $\alpha=0,\ldots,n-1$, labelling the degeneracy of each band in the MBZ. We write the momentum space Hamiltonian in the compact form
	\begin{equation}
		\mathcal{H}=\sum_{\mathbf{k}\in\text{MBZ}}\Psi^\dagger_{{\bf k}} \mathcal{H}_{\text{BdG}}\Psi_{{\bf k}} \, ,
	\end{equation}
	where $\mathcal{H}_{\text{BdG}}$ is the Bogoliubov-de Gennes (BdG) Hamiltonian
	\begin{equation}
		\mathcal{H}_{\text{BdG}}=
		\begin{pmatrix}
			\epsilon({\bf k})-\tilde{\mu} && \Delta\;\mathbb{1}_{n\times n}\\ 
			\Delta^*\;\mathbb{1}_{n\times n} && -\epsilon(-{\bf k})  +  \tilde{\mu}
		\end{pmatrix},
		\label{bdg_Hamiltonian}
	\end{equation}
	with $\epsilon({\bf k})=\text{diag}(\epsilon_0({\bf k}),\ldots,\epsilon_{n-1}({\bf k}))$, and $\tilde{\mu}$ is the shifted chemical potential
	\begin{equation}
		\tilde{\mu}\equiv\mu+\frac{Uf}{2} \, .
		\label{tilde_mu_def}
	\end{equation}

	\begin{figure*}[t!]
		\centering
		\includegraphics[width=0.6\linewidth]{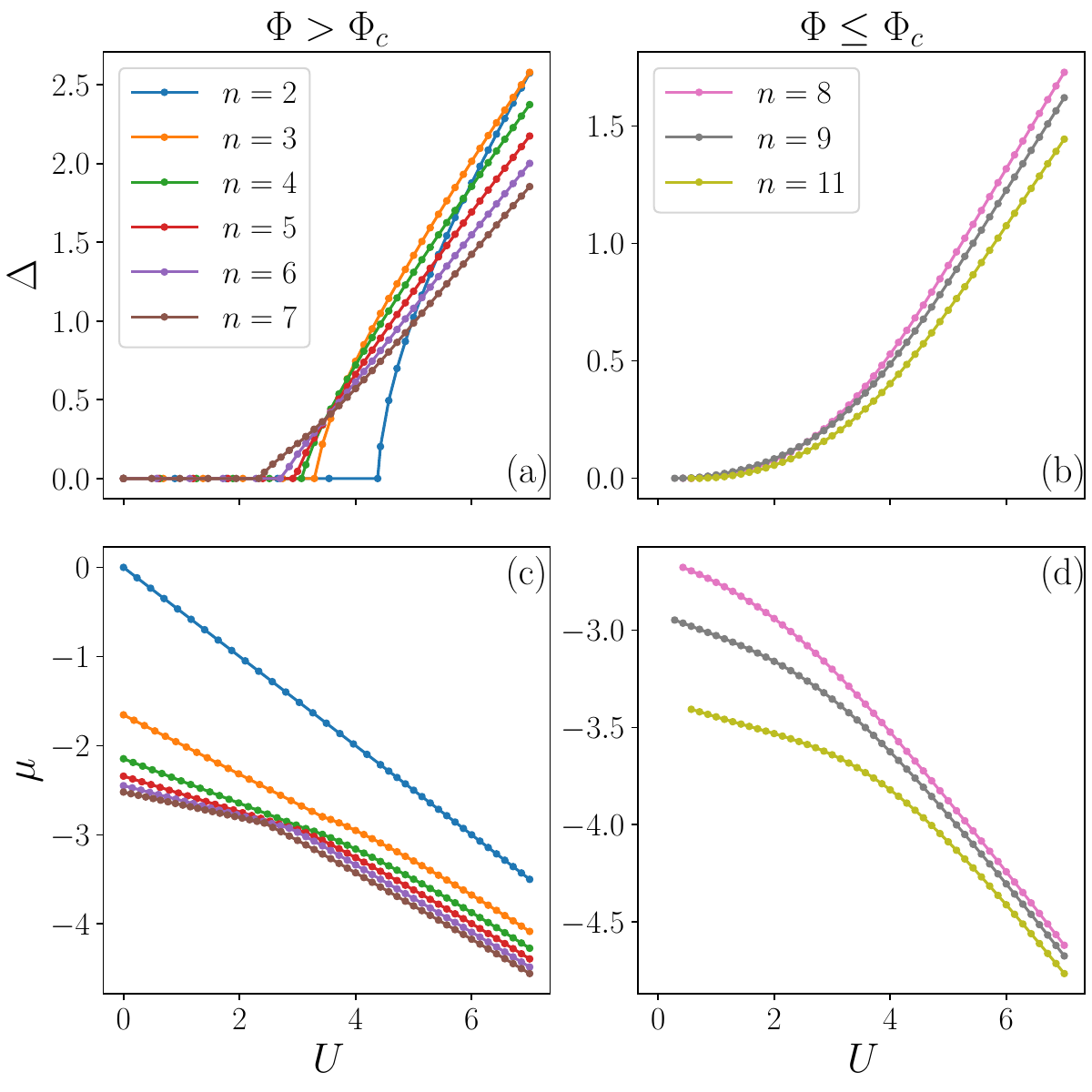}
		\caption{Values of $\Delta$, solutions of the BCS equations as a function of $U$, for fluxes above (left plot) and below (right plot) the critical flux $\Phi_c$. Lines are joining the points to guide the eye for the different trends of the studied coprime pairs.}
		\label{m1_delta_mu_vs_U_linestyle}
	\end{figure*}
    
	We observe that Eq. \eqref{bdg_Hamiltonian} has the same form as the BCS Hamiltonian \cite{Altland_Simons_2010}, with the internal $n\times n$ block structure due to the Hofstadter model in the Hasegawa gauge \cite{IskinPRA2016,IskinPRL2017}. Due to the (both physical and canonical, due to the gauge potential in Eq.\eqref{gauge} \cite{Lepori_Fulga_Trombettoni_BurrelloPRB2016}) inversion symmetry of the magnetic bands, satisfying $\epsilon_\alpha({\bf k})=\epsilon_\alpha(-\bf{k})$, the eigenvalues of the BdG Hamiltonian are $E_{{\bf k}}=\pm\sqrt{\tilde{\epsilon}_{\alpha\mathbf{k}}^2+\Delta^2}$, with $\tilde{\epsilon}_{\alpha\mathbf{k}}\equiv\epsilon_\alpha(\mathbf{k})-\tilde{\mu}$ and the self-consistent BCS equations can be written as
	\begin{equation}
		\frac{1}{U}=\frac{1}{2V}\sum_{\alpha}\sum_{\mathbf{k}\in\text{MBZ}}\frac{1}{\sqrt{\tilde{\epsilon}_{\alpha\mathbf{k}}^2+\Delta^2}} \, ,
		\label{gap_equation}
	\end{equation}
	\begin{equation}
		f=\frac{1}{V}\sum_{\alpha}\sum_{\mathbf{k}\in\text{MBZ}}\bigg[1-\frac{\tilde{\epsilon}_{\alpha\mathbf{k}}}{\sqrt{\tilde{\epsilon}_{\alpha\mathbf{k}}^2+\Delta^2}}\bigg] \, .
		\label{particlenumber_equation}
	\end{equation}
	As for the BdG Hamiltonian, they have the same functional form as the BCS equations, with the additional sum over the magnetic bands. These equations hold at $T=0$, but can be generalized to finite temperatures by minimizing the free energy of the system. Moreover, even if we focused on the simplest scenario of uniform pairing field, analogous equations can be derived for all the possible pairing channels, classifying the Hofstadter SC phases based on the irreducible representations of the magnetic translation group. \cite{SantosPRB2021}.
    
	Finally, we write Eqs. \eqref{gap_equation} and \eqref{particlenumber_equation} by introducing the density of states (DOS) of the system, defined as
	\begin{equation}
		\rho(\epsilon)\equiv\frac{1}{V}\sum_{\mathbf{k}\in\text{MBZ}}\delta(\epsilon-\epsilon(\mathbf{k})) \, .
	\end{equation}
	The sums over the MBZ turn into energy integrals weighted with the DOS, i.e., $V^{-1}\sum_{\alpha,{\bf k}}\rightarrow\int d\epsilon\rho(\epsilon)$, therefore 
	\begin{equation}
		\frac{1}{U}=\frac{1}{2}\int_{-\epsilon_0}^{\epsilon_0}\;d\epsilon\;\frac{\rho(\epsilon)}{\sqrt{\tilde{\epsilon}^2+\Delta^2}} \, ,
		\label{1st_SCE_T0}
	\end{equation}
	\begin{equation}
		f=\int_{-\epsilon_0}^{\epsilon_0}\;d\epsilon\;\rho(\epsilon)\bigg[1-\frac{\tilde{\epsilon}}{\sqrt{\tilde{\epsilon}^2+\Delta^2}}\bigg] \, ,
		\label{2nd_SCE_T0}
	\end{equation}
	where $\epsilon_0$ is the absolute ground state energy of the non-interacting Hamiltonian. This last step allows for the investigation of the SC properties of the HH model based on the energy bands structure and properties. In our analysis we consider $f=2\nu=2m/n$, filling the bands up to the Weyl energy $\epsilon=E_w$ when $\Phi>\Phi_c$, and investigate the critical properties of the model when passing from this regime to the case $\Phi<\Phi_c$.
	
	\section{\label{results_section}Results}
	We numerically solve the coupled self-consistent equations for several coprime pairs $(m,n)$, up to $m=7$, using a Newton algorithm with backtracking, described in detail in Appendix \ref{Newton_method_appendix}. For clarity, in this Section we discuss the representative case $m=1$, for which the critical flux has integer $n_c=7$, as the solutions for other fluxes are qualitatively similar. We first present the numerical results for the pairing field and chemical potential, supported by simple analytical estimates derived from mean-field theory as benchmarks. We then present general arguments, for any value of the flux $\Phi$, characterizing the transition across the critical flux in the plane of the coprime pairs.
	
	\subsection{\label{delta_mu_subsection}Pairing field and chemical potential}
	The results for the pairing field $\Delta$ and the chemical potential $\mu$ are shown in Figs. \ref{m1_delta_mu_vs_U_linestyle}(a), (b) and \ref{m1_delta_mu_vs_U_linestyle}(c), (d). For fluxes $\Phi<\Phi_c$, we observe a quantum phase transition from a semimetallic phase ($U<U_c$) to a superconductive phase ($U>U_c$). The second phase is characterized by a non-zero pairing field $\Delta\neq0$ above a critical interaction $U_c$, which depends on the magnetic flux. As shown in Fig. \ref{m1_delta_mu_vs_U_linestyle}(a), increasing $n$ (i.e., decreasing $\Phi$) leads to a progressive reduction of the critical interaction $U_c$, and, at the same time, modifying the functional behaviour of $\Delta(U)$ in the superconductive phase. When the critical integer $n_c=7$ is crossed, the energy bands overlap and $U_c\rightarrow 0$, with a completely different behavior of $\Delta(U)$, as illustrated in Fig. \ref{m1_delta_mu_vs_U_linestyle}(b). We provide a quantitative analysis of these features in the next Subsection. Interestingly, at fixed $m\geq1$ and for sufficiently large values of $n$, we also observe a crossing of the $\Delta(U)$ curves at an interaction strength $U^*>U_c$. We postulate that this feature could be related to a nontrivial behavior when approaching the thermodynamic limit, where $n\to\infty$ at fixed flux ratio $m/n$, together with $L\to\infty$. A detailed investigation of this aspect, however, goes beyond the scope of the present work and is left for future studies.
	
\begin{figure*}[t!]
		\centering
		\includegraphics[width=0.8\linewidth]{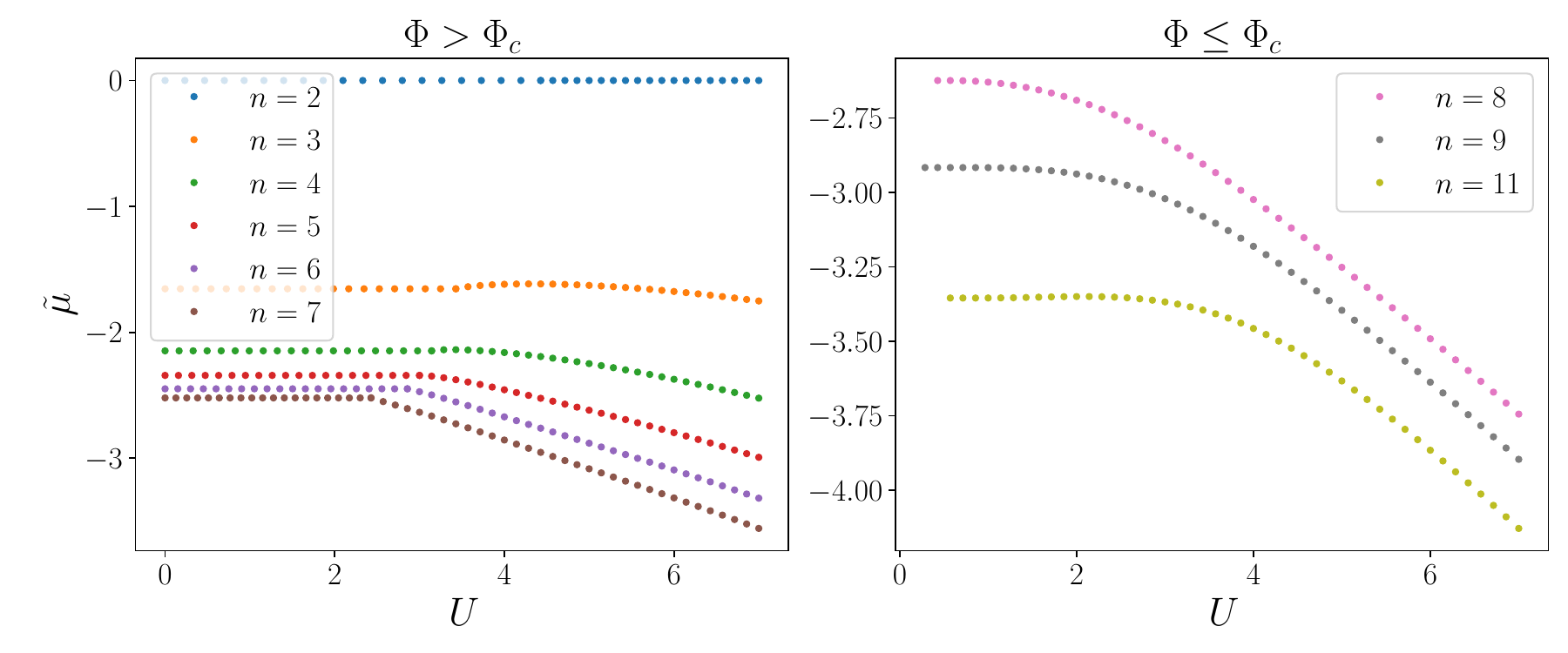}
		\caption{Values of $\tilde{\mu}\equiv\mu+mU/n$ for fluxes above (left plot) and below (right plot) the critical flux $\Phi_c$.}
		\label{m1_tildemu_vs_U_linestyle}
	\end{figure*}

	The chemical potential $\mu$ is a monotonically decreasing function of $U$, regardless of the flux $\Phi$. However, for $\Phi>\Phi_c$, the chemical potential is a linear function of $U$ in the semimetallic phase, $\mu\propto -U$, with small deviations from linearity emerging in the superconductive regime, as shown in Fig. \ref{m1_delta_mu_vs_U_linestyle}(c). This behavior can be better quantified by $\tilde{\mu}$, shown in Fig. \ref{m1_tildemu_vs_U_linestyle}. In the opposite regime, when $\Phi<\Phi_c$, the linear trend is no longer present. 

	\subsection{\label{mu_tilde_MF_computation}Shifted chemical potential in the normal phase}
	The value of the shifted chemical potential $\tilde{\mu}$, defined in Eq. \eqref{tilde_mu_def}, depends on the filling $f=2m/n$, which is fixed by the position of the Weyl nodes in the band structure. At $\Phi=\pi$, the Dirac cones reside exactly at half-filling, i.e., $f=1$. This special case corresponds to a Weyl energy $E_w=0$, and the DOS is symmetric around this point. Due to the particle-hole symmetry, the shifted chemical potential is constant, $\tilde{\mu}=0$, for any interaction strength \cite{Lepori2010}.
	
	For general fluxes, being the Weyl energy $E_w\neq0$, we can no longer apply the symmetry argument. However, we can still derive the value of $\tilde{\mu}$ in the normal phase, where $\Delta=0$ and the self-consistent BCS equations simplify. Specifically, Eq. \eqref{1st_SCE_T0} reduces to 
	\begin{equation}
		1-\frac{2m}{n}=\int_{-\epsilon_0}^{\epsilon_0}\;d\epsilon\;\rho(\epsilon)\;\text{sign}(\epsilon-\tilde{\mu}) \, .
	\end{equation}
	We observe that, setting $\tilde{\mu}=E_w$, this equation yields
	\begin{equation}
		\int_{-\epsilon_0}^{\epsilon_0}\;d\epsilon\;\rho(\epsilon)\;\text{sign}(\epsilon-E_w)=1-\frac{2m}{n},
	\end{equation}
	where we used the normalization of the DOS and the identity
	\begin{equation}
		\frac{m}{n}=\int_{-\epsilon_0}^{E_w}\;d\epsilon\;\rho(\epsilon) \, ,
	\end{equation}
	where $E_w<0$. We conclude that
	\begin{equation}
		\tilde{\mu}=E_w,\qquad\text{for}\;U\leq U_c\;(\Delta=0) \, ,
	\end{equation}
	i.e., in the normal phase $\tilde{\mu}$ is constant and equals the Weyl energy of the lowest node in the spectrum. This result is confirmed by the numerical solutions of the BCS equations, as shown in Fig. \ref{m1_tildemu_vs_U_linestyle}, where we observe the constant value $\tilde{\mu}=E_w$ for $U<U_c$.
	
	\subsection{\label{Uc_MF_computation}Critical value for the interaction strength}
	The result for $\tilde{\mu}$ also allows us to determine the critical interaction strength $U_c$ for $\Phi>\Phi_c$. Indeed, at the transition point, where $U=U_c$, we have $\Delta=0$ and $\tilde{\mu}=E_w$, then Eq. \eqref{1st_SCE_T0} becomes
	\begin{equation}
		\frac{1}{U_c}=\frac{1}{2}\int_{-\epsilon_0}^{\epsilon_0}\;d\epsilon\;\frac{\rho(\epsilon)}{|\epsilon-E_w|} \, .
		\label{Uc_equation}
	\end{equation}
	In Table \ref{table:Uc_values_m1} we report the values of $U_c$ estimated for various integers at fixed $m=1$, together with their error, associated with the numerical precision in the evaluation of the integral. The predicted values agree with the results shown in Fig. \ref{m1_delta_mu_vs_U_linestyle}. 
	
	\begin{table}[t]
		\setlength{\tabcolsep}{10pt}
		\centering
		\begin{tabular}{c|cc}
			\hline
			\hline
			$m$ & $n$ & $U_c$ \\ 
			\hline
            1 & 2 & 4.382(2) \\
			  & 3 & 3.29(3) \\ 
			& 4 & 3.08(3) \\ 
			& 5 & 2.9(1) \\ 
			& 6 & 2.74(7) \\ 
			& 7 & 2.2(2) \\ 
			\hline
			\hline
		\end{tabular}
		\caption{Values of the critical interaction $U_c$ for several coprime pairs $(1,n)$. For the pair $(1,2)$, corresponding to the $\pi$-flux, the result is already available in the literature \cite{Lepori2010} and agrees with Eq. \eqref{Uc_equation}, while for $n>n_c=7$ no phase transition occurs due to band overlap.}
		\label{table:Uc_values_m1}
	\end{table}
	
	\subsection{\label{final_remarks_results}Transition across the critical flux line}
	The features discussed for the representative case $m=1$ extend to all coprime pairs $(m,n)$. For each of these, a critical interaction strength $U_c$ separates the semimetallic and superconductive phases when $\Phi>\Phi_c$. The full phase diagram in the $(m,n)$ plane, shown in Fig. \ref{n_vs_m_phasediagram_Uc}, reveals a critical line of slope $1/\Phi_c$, which separates the region where $U_c\neq0$, identifying a quantum phase transition with well-defined Weyl nodes in the energy spectrum, from the region where $U_c=0$, characterized by a full bands overlap and screened Weyl points \cite{FontanaPRB2021}. Points lying on this critical line correspond to the limiting case of $\Phi=\Phi_c$.
	
	\section{\label{scaling_Uc_phi_subsection}Scaling relations and critical exponents}
	Having presented the numerical solutions to the BCS equations in Eqs. \eqref{1st_SCE_T0} and \eqref{2nd_SCE_T0} for the pairing field $\Delta$ and the chemical potential $\mu$ across the semimetal to superconductor transition, we now focus on their critical properties close to the transition point. In this Section, we analyze the emergence of scaling behavior close to $U\simeq U_c$ in the neighborhood of the critical line $\Phi_c^{-1}$ shown in Fig. \ref{n_vs_m_phasediagram_Uc}, approaching the transition from below and above. 
    
    \subsection{\label{scaling_large_fluxes} Below the critical flux line $\Phi_c^{-1}$}
	When $U_c\neq0$, i.e., for fluxes $\Phi>\Phi_c$, we focus on the scaling behavior of the pairing field and the shifted chemical potential. 
	\begin{equation}
		\Delta=\Delta_0\, (U-U_c)^{\beta(\Phi)},\qquad \Delta_0>0 \, ,
		\label{beta_phi_def}
	\end{equation}
    \begin{equation}
		\tilde{\mu}-E_w=\mu_0 \, (U-U_c)^{\alpha(\Phi)} \, ,
		\label{alpha_phi_def}
	\end{equation}
    where the exponents $\alpha(\Phi)$ and $\beta(\Phi)$, as well as the sign of $\mu_0$, may in principle depend on the magnetic flux.

    Combining analytical and numerical arguments, we find that the value of the scaling exponents is universal, i.e., independent of the magnetic flux, with the leading MF exponent for the pairing field given by $\beta=1/2$ (see Appendix \ref{pairing_field_scaling} for the analytical derivation from the gap equation, including possible marginal corrections) and the one for the shifted chemical potential given by $\alpha=1$. 

    \begin{figure}[t]
        \centering
        \includegraphics[width=1.0\linewidth]{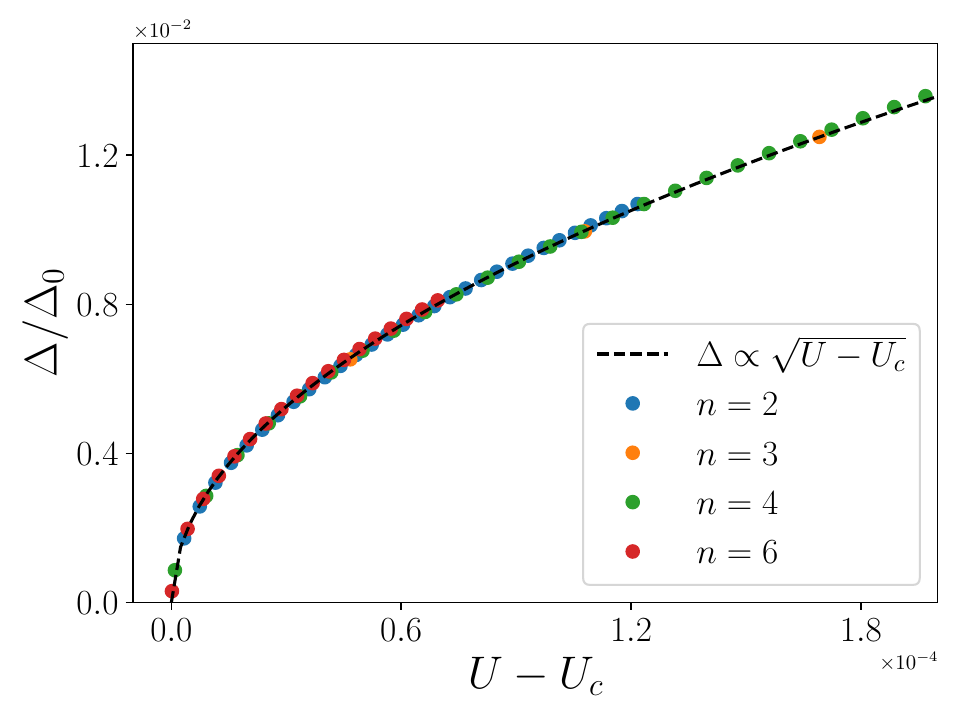}
        \caption{Scaling plot of $\Delta/\Delta_0$, as a function of $U-U_c$, for all values discussed in Appendix \ref{Hasegawa_implicit_relations}, and at $m=1$. We plot as well the function $\Delta\propto \sqrt{U-U_c}$ (black dashed line) for a direct comparison. The data for different magnetic fluxes collapse very well, confirming the universality of the mean-field exponent $\beta(\Phi) = \beta = 1/2$.}
        \label{fig:scaling_delta_vs_U-Uc_Hasegawa_n}
    \end{figure}

    \begin{figure}[t]
		\centering
		\includegraphics[width=1.0\linewidth]{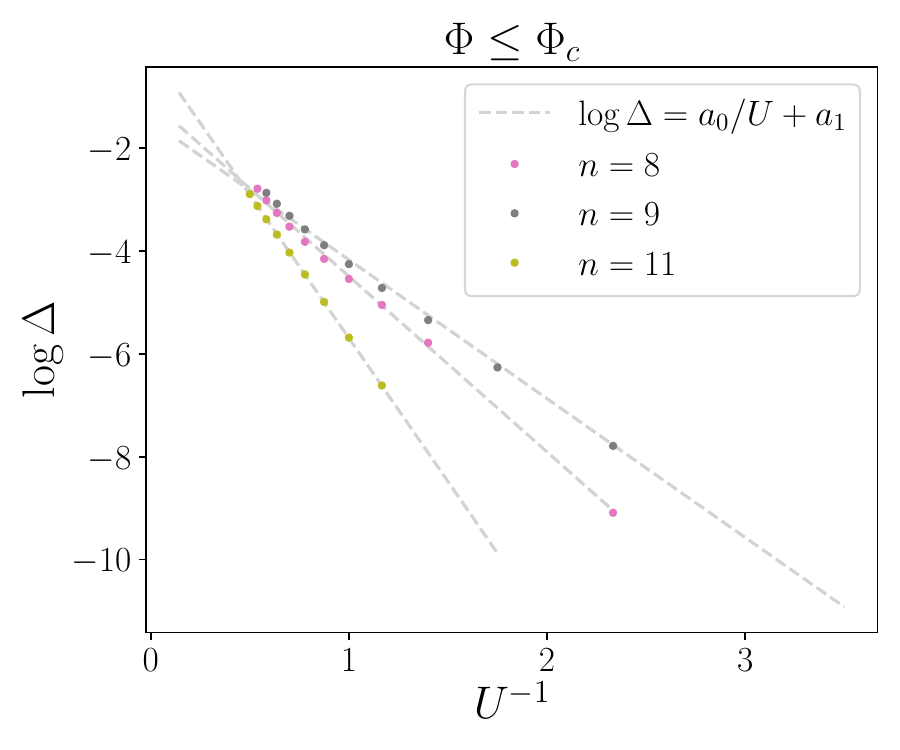}
		\caption{Logarithm of the pairing field $\Delta$ as a function of $U^{-1}$ for different $n>n_c=7$ at fixed $m=1$. The dashed grey lines correspond to the fit in Eq. \eqref{exponential_closure_delta} with the log-scale on the vertical axis, of the form $\log\Delta=a_0/U+a_1$.}
		\label{logDelta_vs_Um1_fit}
	\end{figure}

    \begin{figure*}
        \centering
        \includegraphics[width=1.0\linewidth]{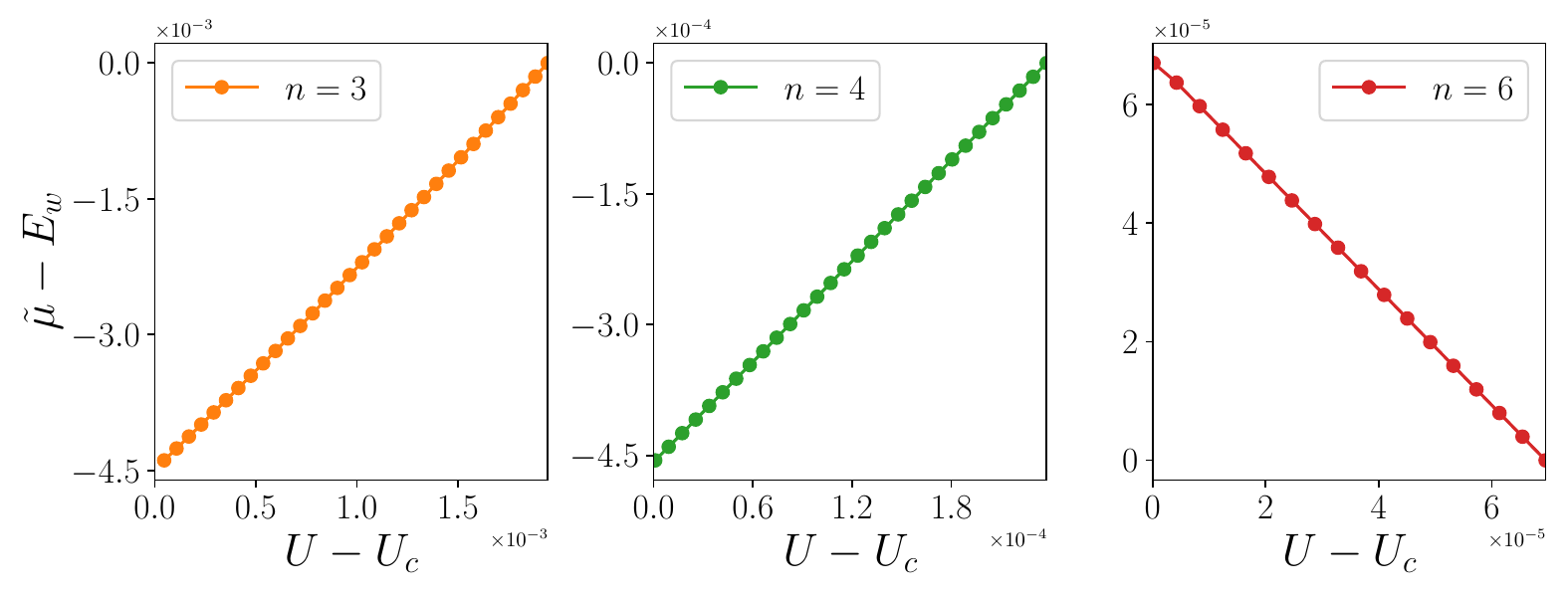}
        \caption{Plots of $\tilde{\mu}-E_w$, as a function of $U-U_c$ and for the values discussed in Appendix \ref{Hasegawa_implicit_relations}, and at $m=1$. The special case $n=2$ (the $\pi$-flux) is excluded, since $\tilde{\mu}=0$ due to particle-hole symmetry at the Weyl filling. For $n\ge 3$, the data show linear scaling with exponent $\alpha(\Phi) = \alpha = 1$, with the sign of $\mu_0$ depending on $n$, as evident from the right panel.}
        \label{fig:tilde_mu_vs_U-Uc_Hasegawa_n}
    \end{figure*}
    
    This universal scaling can be obscured when solving the self-consistent equations \eqref{gap_equation} and \eqref{particlenumber_equation} at limited numerical precision, e.g., in the numerical sampling of the DOS or a coarse sampling of $U$. To circumvent possible numerical inaccuracies, we complement the results of Section \ref{results_section} with computations in momentum space based on the implicit exact relations derived in Ref. \cite{Hasegawa1990} for specific $n$ values (see Appendix \ref{Hasegawa_implicit_relations}), which facilitate the sampling around the critical point. This allows us to observe that, when restricted to $U\simeq U_c$, the scaling of the pairing field is not only universal but also with $\beta=1/2$. This is clear in Fig. \ref{fig:scaling_delta_vs_U-Uc_Hasegawa_n}, where the curves for different $n$ collapse onto each other according to Eq. \eqref{beta_phi_def} with $\beta(\Phi)=\beta=1/2$. 
    
    Using the same technique for the shifted chemical potential $\tilde{\mu}-E_w$, we observe that, apart from $n=2$, where particle-hole symmetry at half-filling enforces $\tilde{\mu}=0$ \cite{Mazzucchi_2013}, for $n\geq3$ there is a linear scaling, i.e., $\alpha(\Phi)=\alpha=1$, with the sign of $\mu_0$ in Eq. \eqref{alpha_phi_def} possibly depending on $n$, as depicted in Fig. \ref{fig:tilde_mu_vs_U-Uc_Hasegawa_n}.

    \subsection{\label{scaling_small_fluxes}Above the critical flux line $\Phi_c^{-1}$}
	When $\Phi<\Phi_c$ the model does not have a critical interaction strength (see Fig. \ref{n_vs_m_phasediagram_Uc}), and $\Delta\rightarrow0$ for $U\rightarrow0$. 
    Since the filling $f=2m/n$ in this case correspond to having $(m+1)$ partially occupied bands \cite{FontanaPRB2021}, we expect an exponential closure of the gap with the inverse of the interaction strength \cite{IskinPRL2017}, according to the scaling 
	\begin{equation}
		\Delta=\Delta_1\exp{\bigg(-\frac{b}{\rho(E_f)U}\bigg)} \, ,
		\label{exponential_closure_delta}
	\end{equation}
	where $\Delta_1$ and $b$ are free parameters, and $\rho(E_f)$ is the DOS corresponding to the energy $E_f$ such that
	\begin{equation}
		\int_{-\epsilon_0}^{E_f}\rho(\epsilon) \, d\epsilon=\frac{m}{n} \, .
	\end{equation}
	The data and fit curves for the representative case of $m=1$ are shown in Fig. \ref{logDelta_vs_Um1_fit}, highlighting the exponential closure of the gap in the $(U^{-1},\log\Delta)$ plane. The fit parameters $\Delta_1$ and $b$ of Eq. \eqref{exponential_closure_delta} relate to $a_0$ and $a_1$, introduced in Fig. \ref{logDelta_vs_Um1_fit}, as $a_0=-b/\rho(E_f)$ and $a_1=\log\Delta_1$, and their values are reported in Table \ref{table:fit_params_exp_closure}.

    \begin{table}[t]
		\setlength{\tabcolsep}{10pt}
		\centering
		\begin{tabular}{c|ccc}
			\hline
			\hline
			$m$ & $n$ & $a_0$ & $a_1$ \\ 
			\hline
			1 & 8 & -3.42(4) & -1.07(5) \\ 
			& 9 & -2.70(3) & -1.47(5)\\ 
			& 11 & -5.57(1) & -0.12(1)\\
			\hline
			\hline
		\end{tabular}
		\caption{Fit parameters for the function $\log\Delta=a_0/U+a_1$, with $a_0=-b/\rho(E_f)$ and $a_1=\log\Delta_1$ to relate it with Eq. \eqref{exponential_closure_delta}.}
		\label{table:fit_params_exp_closure}
	\end{table}
    
	\subsection{\label{Uc_vs_phi_scaling}Critical interaction strength}
	\begin{figure}[t]
		\centering
		\includegraphics[width=1.0\linewidth]{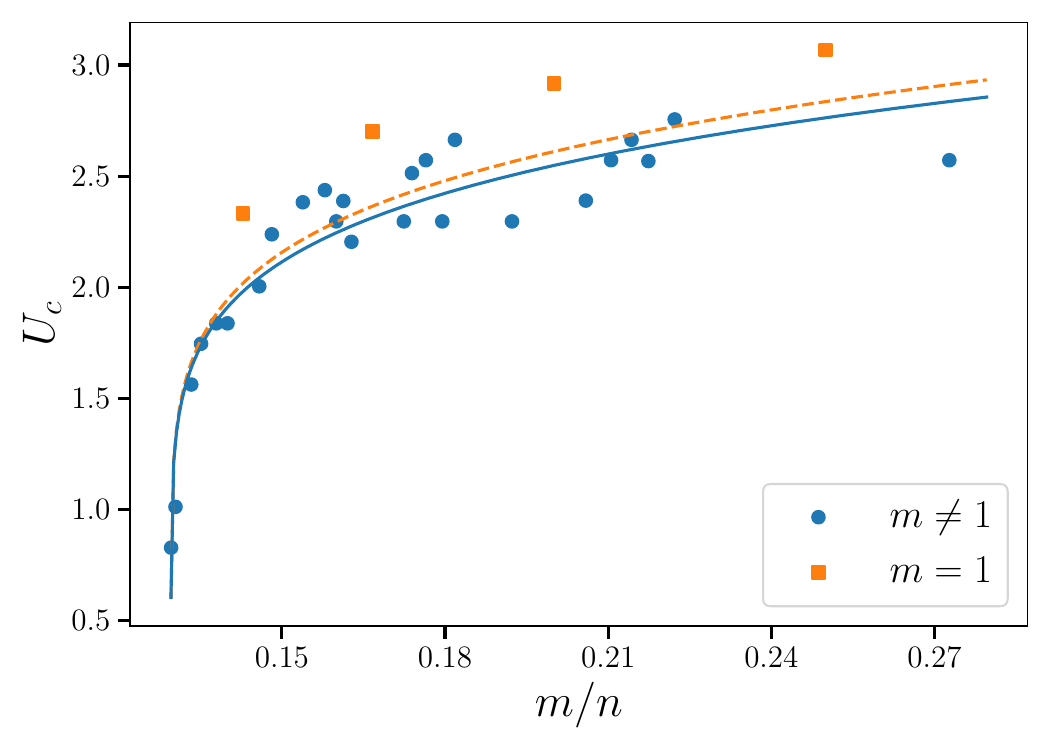}
		\caption{Critical interaction $U_c$ as function of $m/n=\Phi/2\pi$ for $m=1$ (orange markers) and $m\neq1$ (blue markers), with the best fits of the scaling function in Eq. \eqref{Uc_vs_phi_gamma_def}, respecting the same color code.}
		\label{Uc_vs_movern_with_fits}
	\end{figure}
	As a final point, we investigate the dependence of $U_c$ on the magnetic flux for arbitrary coprime pairs $(m,n)$. Based on the phase diagram in Fig. \ref{n_vs_m_phasediagram_Uc}, we expect that the critical strength $U_c$ goes to zero as we approach $\Phi_c$ from above, i.e., $\Phi\rightarrow\Phi_c^+$. For different vertical cuts of the phase diagram, i.e. at fixed values of $m$, we extract $U_c$ and progressively increase $n$ to cross the critical line. The observed behaviors are consistent with the law $U_c\sim(n_c-n)^{\tilde{\beta}}$, with $\tilde{\beta}>0$ and $n_c=n_c(m)$ being the critical integer associated to $m$ \cite{FontanaPRB2024}. However, in view of extracting the behavior for $\Phi\rightarrow\Phi_c$, we expect that the SC properties of the model depends on the magnetic field itself rather than its explicit parametrization as a ratio of coprime integers. We therefore investigate the dependence of $U_c$ on the ratio $m/n$, with the scaling ansatz
	\begin{equation}
		U_c=U_0\bigg(\frac{m}{n}-\frac{m}{n_c}\bigg)^\gamma\sim(\Phi-\Phi_c)^\gamma \, ,\qquad \gamma>0 \, .
		\label{Uc_vs_phi_gamma_def}
	\end{equation}
	The data and fit functions, with free parameters $U_0,\;\gamma$ and $\Phi_c/2\pi$ are plotted in Fig. \ref{Uc_vs_movern_with_fits}. We graphically separate the case of $m=1$, treated as representative case, from all the other integers for visual purposes. The scaling functions differ for $\Phi>\Phi_c$, as highlighted in the plot, but the best fit values for $\gamma$ are compatible (we refer to Appendix \ref{stability_fit_parameters} for a detailed analysis of the convergence of the fit parameters). The estimates of the fit parameters are $U_0=3.8(2)$ and $\gamma=0.15(1)\approx 3/20$, with a critical flux $\Phi_c/2\pi=0.1296(1)$, in agreement with the conjectured asymptotic value \cite{FontanaPRB2024}.
    
	\section{\label{conclusions}Conclusions}
    In this work, we analyze the superconducting (SC) properties of the three-dimensional Hofstadter--Hubbard (HH) model below the critical rational flux where Weyl points appear in the spectrum. In particular, we explore the mean-field (MF) phase diagram for varying rational fluxes parametrized as coprime ratios, and perform a scaling analysis around the quantum critical points.

    Our results show that the SC features of the model are strongly controlled by the underlying magnetic band structure and, in particular, by the presence of the critical flux $\Phi_c$ separating two distinct regimes. For large fluxes, i.e., $\Phi>\Phi_c$, where the single-particle spectrum exhibits Weyl nodes, we find a semimetal-to-superconductor quantum phase transition occurring at a finite critical interaction strength ($U_c\neq0$). In this regime, the pairing field vanishes continuously at the transition and displays universal MF scaling, with $\Delta\sim (U-U_c)^{1/2}$, while the shifted chemical potential shows linear behaviour, i.e. $\tilde{\mu}-E_w\sim U-U_c$. In contrast, for small fluxes, i.e., $\Phi<\Phi_c$, where the magnetic bands overlap and no isolated zeros are present in the density of states, SC emerges for arbitrarily small attractive interactions. Here the pairing field exhibits an exponential dependence on $U^{-1}$, consistent with standard BCS behaviour \cite{Grosso_Parravicini_2000,El-Batanouny2020}. These two regimes are separated by the critical flux line $\Phi=\Phi_c$, which thus acts as a boundary between qualitatively different SC responses.

    We further determine the dependence of the critical interaction $U_c$ on the magnetic flux, showing that it vanishes continuously as $\Phi\to\Phi_c^+$ according to a power law scaling in $\Phi-\Phi_c$, providing a unified description of the phase diagram in terms of the coprime flux ratio. 
    
    As a possible future development, we mention a field-theoretical description close to the critical line  $\Phi_c$ and for $m\rightarrow\infty$, to complement our present results based on the numerical solutions of the self-consistent equations. Other intertwined directions involve the finite-temperature regime with corrections from fluctuations \cite{Park2020}, as well as the inspection of flat band features emerging in the regime of small magnetic fluxes \cite{Peotta2015}.
	
	\section{Acknowledgements}
	We are very thankful to P. M. Bonetti, M. Burrello and M. B. Silva Neto for discussions and suggestions. L. L. acknowledges funding from the European Union—NextGenerationEU, PNRR MUR Project PE0000023-NQSTI. 
	
	\bibliography{biblio}
	
	\appendix
	\widetext
	\section{\label{Hofstadter_model_appendix}Diagonalization of the $3D$ Hofstadter model}
	We present here a sketch of the method used to diagonalize $\mathcal{H}_{\text{Hofst}}$ in momentum space, referring to \cite{BurrelloJPHYSMATH2017,FontanaPRB2024} for a review of the properties of this Hamiltonian. Due to the presence of the commensurate field, the cubic lattice can be decomposed in sub-lattice whose size depends on the choice of the gauge. This gauge freedom can be exploited to select a gauge that minimizes the number of sub-lattices associated to the magnetic flux $\Phi$, and allows for the diagonalization of the problem with the minimum possible effort. It has been proved that, within this setting, the minimal number of independent sub-lattices is precisely $n$ \cite{BurrelloJPHYSMATH2017}, and a convenient gauge choice is the Hasegawa gauge \cite{Hasegawa1990}, defined through the vector potential 
	\begin{equation}
		\bm{A}(\mathbf{x})=\Phi(0,x-y,y-x).
		\label{hasegawa_gauge}
	\end{equation}
	The relative Peierls phases are
	\begin{equation}    
		\theta_x(\bm{r})=0,\qquad\theta_y(\bm{r})=\Phi\bigg(x-y-\frac{1}{2}\bigg),\qquad\theta_z(\bm{r})=\Phi(y-x) \, .
		\label{peierls_phases}
	\end{equation}
	A notable feature emerging from Eq. \eqref{peierls_phases} is its explicit dependence on the relative coordinate $(x-y)$, and the absence of the $z$-coordinate, indicating that the problem is effectively one-dimensional in reciprocal space \cite{KunsztZeePRB1991}.
	
	The free Hamiltonian $\mathcal{H}_{\text{Hofst}}$ can be efficiently diagonalized in momentum space, introducing the magnetic Brillouin zone (MBZ)
	\begin{equation}
		\text{MBZ}:\quad k_x\in\bigg[-\frac{\pi}{n} \, ,\frac{\pi}{n}\bigg],\;k_y\in\bigg[-\frac{\pi}{n},\frac{\pi}{n}\bigg] \, ,\;k_z\in\bigg[-\pi,\pi\bigg] \, ,
		\label{MBZ_definition}
	\end{equation}
	that exploits the interplay between gauge and translational invariance in the presence of a commensurate magnetic flux. The Hamiltonian is decomposed into independent blocks, corresponding to degenerate magnetic bands associated with distinct sub-lattices.
	
	The Hamiltonian in momentum space takes the form 
	\begin{equation}
		H=-t\sum_{\mathbf{k}\in\text{MBZ}}\sum_{\hat{j},s}c^\dag_{s',\mathbf{k}}(T_{\hat{j}})_{s',s}e^{-i\mathbf{k}\cdot\hat{j}}c_{s,\mathbf{k}}+\text{H.c.}\equiv-t\sum_{\mathbf{k}\in\text{MBZ}}C^\dagger_{\mathbf{k}}\mathcal{H}(\mathbf{k})C_{\mathbf{k}}\,,
		\label{reduced_H_momentumspace}
	\end{equation}
	where we defined the vector of operators $C^\dagger_{{\bf k}}=(c^\dagger_{0{\bf k}},\ldots,c^\dagger_{{n-1},{\bf k}})$, $s$ is the label for the magnetic bands and the $n \times n$ matrices $T_{\hat{j}}$, with $\hat{j}=\hat{x},\hat{y},\hat{z}$, are defined as 
	\begin{equation}
		T_{\hat{x}}=
		\begin{pmatrix}
			0 & 1 & 0 & 0\\
			0 & 0 & \ddots & 0\\
			0 & \cdots & 0 & 1\\
			1 & 0 & \cdots & 0
		\end{pmatrix},
		\qquad
		T_{\hat{y}}=e^{-\frac{i\Phi}{2}}
		\begin{pmatrix}
			0 & \cdots & 0 & \varphi_0\\
			\varphi_1 & 0 & \cdots & 0\\
			0 & \ddots & 0 & 0 \\
			0 & 0 & \varphi_{n-1} & 0
		\end{pmatrix},
		\qquad
		T_{\hat{z}}=
		\begin{pmatrix}
			\varphi_0 & 0 & \cdots & 0\\
			0 & \varphi_{n-1} & 0 & 0\\
			0 & 0 & \ddots & 0\\
			0 & \cdots & 0 & \varphi_1
		\end{pmatrix} \, .
		\label{Tmatrices_defs}
	\end{equation} 
	In Eq. \eqref{Tmatrices_defs} we introduced $\varphi_l=e^{i\Phi l}=e^{\frac{2\pi i m l}{n}}$, with $l=0,\ldots,n-1$. The role of $\Phi$ as commensurate field is of key importance, and is now explicit in the definitions of Eq. \eqref{Tmatrices_defs}. For $n=2$, the system realizes the well-known $\pi$-flux model, hosting a Weyl semimetallic phase at half-filling, preserving both physical time-reversal and space inversion symmetries \cite{Lepori_Fulga_Trombettoni_BurrelloPRB2016}. For $n\neq2$ the time-reversal symmetry is explicitly broken. Finally, under the transformation $c_{\mathbf{r}}\rightarrow(-1)^{x+y+z}c_{\mathbf{r}}$, the real-space Hamiltonian is transformed as $\mathcal{H}\rightarrow-\mathcal{H}$. As a consequence of this chiral sub-lattice transformation, the system has a symmetric single-particle spectrum, reflecting into the relation $\rho(\epsilon)=\rho(-\epsilon)$ for the DOS at generic fluxes.

    \section{\label{Hasegawa_implicit_relations}Implicit relations for the spectra at $m=1$}
    In this Appendix we collect the implicit relations defining the secular equation of the momentum-space Hamiltonian in the Hasegawa gauge, derived in Ref. \cite{Hasegawa1990}. These results specific to the cases of $m=1$ and $n=2,3,4,6$. For these coprime pairs the characteristic polynomial can be written in a closed form, which enables an efficient numerical evaluation of the energy bands without performing the reduced ED discussed in Appendix \ref{Hofstadter_model_appendix}.

    These relations are particularly useful for  
    the numerical solution of the self-consistent BCS equations \eqref{gap_equation} and \eqref{particlenumber_equation}. The spectrum $\epsilon_\alpha({\bf k})$ is obtained directly as the set of real roots of the corresponding polynomial equations. Denoting by $E$ the eigenvalues of the problem $\text{det}[\mathcal{H}({\bf k})-E]=0$, the characteristic polynomials take the following form.
    \begin{itemize}
        \item $(m,n)=(1,2)$, corresponding to $\Phi = \pi$:
        \begin{equation}
            E^2-4t^2(\cos^2{k_x}+\cos^2{k_y}+\cos^2{k_z})=0.
        \end{equation}
        \item $(m,n)=(1,3)$, corresponding to $\Phi = 2\pi/3$:
        \begin{equation}
            -E^3+9t^2E+2t^3(\cos{3k_z}-\cos{3k_x}-\cos{3k_y})-6t^3\cos(k_x+k_y+k_z)=0.
        \end{equation}
        \item $(m,n)=(1,4)$, corresponding to $\Phi = \pi/2$:
        \begin{equation}
            E^4 - 12t^2E^2+8\sqrt{2}t^3E\cos(k_x+k_y+k_z)+2t^4(3-\cos{4k_x}-\cos{4k_y}-\cos{4k_z})=0.
        \end{equation}
        \item $(m,n)=(1,6)$, corresponding to $\Phi = \pi/3$:
        \begin{align}
            \nonumber
            &E^6-18t^2E^4+12\sqrt{3}t^3E^3\cos(k_x+k_y+k_z)+45t^4E^2\\
            &-36\sqrt{3}t^5E\cos(k_x+k_y+k_z)+2t^6[3\cos2(k_x+k_y+k_z)-(\cos 6k_x+\cos 6k_y+\cos 6k_z)]=0.
        \end{align}
    \end{itemize}
    For each ${\bf k}$-value, the spectrum $\epsilon_\alpha({\bf k})$ is obtained. In practice, solving these implicit relations numerically provides substantially faster access to the spectrum than the reduced ED, while remaining fully equivalent at the level of the single-particle problem.
    
	\section{\label{Newton_method_appendix}The Newton method with backtracking}
	We recast the coupled self-consistent equations \eqref{1st_SCE_T0}, \eqref{2nd_SCE_T0} as the system
	\begin{equation}
		{\bf F} = 
		\begin{pmatrix}
			F_1({\bf x})\\
			F_2({\bf x})
		\end{pmatrix}=0 \, ,
		\qquad 
		(\Delta,\mu)\equiv{\bf x} \, ,
	\end{equation}
	where
	\begin{equation}
		F_1(\Delta,\mu)\equiv\frac{1}{U}-\frac{1}{2}\int_{-\epsilon_0}^{\epsilon_0}\;d\epsilon\;\frac{\rho(\epsilon)}{\sqrt{\tilde{\epsilon}^2+\Delta^2}} \, ,\qquad
		F_2(\Delta,\mu)\equiv f-\int_{-\epsilon_0}^{\epsilon_0}\;d\epsilon\;\rho(\epsilon)\bigg[1-\frac{\tilde{\epsilon}}{\sqrt{\tilde{\epsilon}^2+\Delta^2}} \bigg] \, .
	\end{equation}
	We apply the Newton algorithm to solve the system of equations, using the following steps:
	\begin{enumerate}
		\item propose an initial guess ($k=0$) for $(\Delta,\mu)=(\Delta_0,\mu_0)$;
		\item compute, for $k\geq0$:
		\begin{equation}
			{\bf J}({\bf x}_k)\underbrace{({\bf x}_{k+1}-{\bf x}_k)}_{\equiv {\bf s}_s}=-{\bf F}({\bf x}_k) \, ,
		\end{equation}
		where ${\bf J}$ is the Jacobian matrix computed in ${\bf x}_k$. The necessary condition for this step is that the Jacobian matrix must be invertible;
		\item define ${\bf x}_{k+1}={\bf s}_k+{\bf x}_k$, $k\rightarrow k+1$;
		\item repeat until convergence, by fixing a threshold $\delta{\bf x}$ on the solutions, together with a maximum number of iterations to prevent infinite loops. As exit test, we select the relative difference
		\begin{equation}
			||{\bf s}||_k=||{\bf x}_{k+1}-{\bf x}_k||<\delta{\bf x}.
		\end{equation}
	\end{enumerate}
	There could be cases where convergence is not reached, and continuous oscillations around a constant values are present. This can drastically affect the determination of the solution, and the associated precision. This happens in the normal phase, where $\Delta=0$, and the Newton algorithm starts to oscillate between positive and negative values of $\Delta$ around zero, not converging within a finite (even if large) number of steps.
	
	We thus adapt the Newton method with further checks, implementing the \textit{backtracking}, to avoid divergences, in combination with the \textit{Armijo condition}, to avoid oscillations \cite{Armijo1966,Dennis_Review_1977}. The adapted method reads:
	\begin{enumerate}
		\item propose an initial guess ($k=0$) for $(\Delta,\mu)=(\Delta_0,\mu_0)$;
		\item compute, for $k\geq0$:
		\begin{equation}
			{\bf J}({\bf x}_k)\underbrace{({\bf x}_{k+1}-{\bf x}_k)}_{\equiv {\bf s}}=-{\bf F}({\bf x}_k) \, ,
		\end{equation}
		\item implement backtracking and Armijo condition, by proceeding in the Newton direction with
		\begin{equation}
			{\bf x}_{k+1}={\bf x}_k+\alpha{\bf s}_k,
		\end{equation}
		with $\alpha$ the largest value for which there is a sufficient decrease in ${\bf F}({\bf x})$, i.e.
		\begin{equation}
			||{\bf F}({\bf x}_{k+1})||<(1-\tau\alpha)\;||{\bf F}({\bf x}_k)||
			\label{armijo_rule}
		\end{equation}
		and $\tau=10^{-5}$ is a small parameter. Practically, we check Eq. \eqref{armijo_rule} for the initial value of $\alpha=\alpha_0=1$, then:
		\begin{itemize}
			\item if Eq. \eqref{armijo_rule} is satisfied with the proposed value of $\alpha$, accept the Newton move;
			\item if Eq. \eqref{armijo_rule} is not satisfied, we rescale $\alpha\rightarrow \beta\alpha$ by a factor $\beta<1$, that we fix to $\beta=1/2$, and go back to 3.;
		\end{itemize}
		\item the exit test for the whole loop is now provided by the Eq. \eqref{armijo_rule} itself, combined with the condition $||{\bf s}||_k<\delta{\bf x}$ for the precision of the Newton solution. 
	\end{enumerate}
	
	As a final comment, we observe that when we set $\tau=0$ we check that the solutions are not diverging, avoiding explosions of their values between consecutive steps. This is what is usually referred to as the Newton method with backtracking. With the introduction of $\tau$, provided it is small enough, we implement the Armijo condition, to check and prevent oscillations of the solutions.
	
	\subsection{Jacobian evaluation}
	We evaluate here the Jacobian of the vector ${\bf F}(\Delta,\mu)$:
	\begin{equation}
		\frac{\partial F_1}{\partial \Delta}=\frac{\Delta}{2}\int_{-\epsilon_0}^{\epsilon_0}\;d\epsilon\;\frac{\rho(\epsilon)}{(\tilde{\epsilon}^2+\Delta^2)^{3/2}} \, ,\qquad  \frac{\partial F_1}{\partial \mu}=-\frac{1}{2}\int_{-\epsilon_0}^{\epsilon_0}\;d\epsilon\;\frac{\tilde{\epsilon}\;\rho(\epsilon)}{(\tilde{\epsilon}^2+\Delta^2)^{3/2}} \, ,
		\label{dF1_Jacobian_computation}
	\end{equation}
	\begin{equation}
		\frac{\partial F_2}{\partial \Delta}=-\Delta\int_{-\epsilon_0}^{\epsilon_0}\;d\epsilon\;\frac{\tilde{\epsilon}\;\rho(\epsilon)}{(\tilde{\epsilon}^2+\Delta^2)^{3/2}} \, ,\qquad  \frac{\partial F_1}{\partial \mu}=-\Delta^2\int_{-\epsilon_0}^{\epsilon_0}\;d\epsilon\;\frac{\rho(\epsilon)}{(\tilde{\epsilon}^2+\Delta^2)^{3/2}}.
	\end{equation}
	We observe the relations
	\begin{equation}
		\frac{\partial F_2}{\partial \Delta}=2\Delta\frac{\partial F_1}{\partial \mu} \, ,\qquad\qquad \frac{\partial F_2}{\partial \mu}=-2\Delta\frac{\partial F_1}{\partial \Delta} \, ,
	\end{equation}
	essentially reducing the computation of the Jacobian to the derivatives of $F_1$. For completeness, the explicit matrix expression is
	\begin{equation}
		{\bf J}(\Delta,\mu)=
		\begin{pmatrix}
			\frac{\partial F_1}{\partial \Delta} & \frac{\partial F_1}{\partial \mu}\\\\
			2\Delta \frac{\partial F_1}{\partial \mu} & -2\Delta \frac{\partial F_1}{\partial \Delta}
		\end{pmatrix}.
	\end{equation}
	
	\section{\label{mneq1_results}Numerical results for coprime pairs $(m,n)$ with $m\leq 7$}
    
    We discuss the numerical results for $m>1$, complementing the analysis presented in Section \ref{results_section} for the reference case $m=1$. 
    \begin{table}[t]
		\setlength{\tabcolsep}{10pt}
		\centering
		\begin{tabular}{c|cccc}
			\hline
			\hline
			$m$ & $U_0$ & $n_c$ & $\tilde{\beta}$ & $R^2$ \\
			\hline
			1 & 2.2(3) & 8.3(6)  & 0.23(6) & 0.995 \\
			2 & 2.3(1) & 15.01(4) & 0.09(4) & 0.978 \\
			3 & 2.3(1) & 23 & 0.05(3) & 0.992 \\
			4 & 2.0(1) & 29.4(4) & 0.10(3) & 0.976 \\
			5 & 2.0(5) & 37.1(9) & 0.07(9) & 0.905 \\
			7 & 1.49(3) & 54.03(1) & 0.160(8) & 0.999 \\
			\hline
			\hline
		\end{tabular}
		\caption{Power-law fit parameters to Eq. \eqref{eq:Uc_various m_fit} for the considered values of $m$.}
		\label{table:powerlaw_fit_results}
	\end{table}

    \begin{figure}[t]
		\centering
		\includegraphics[width=1.0\linewidth]{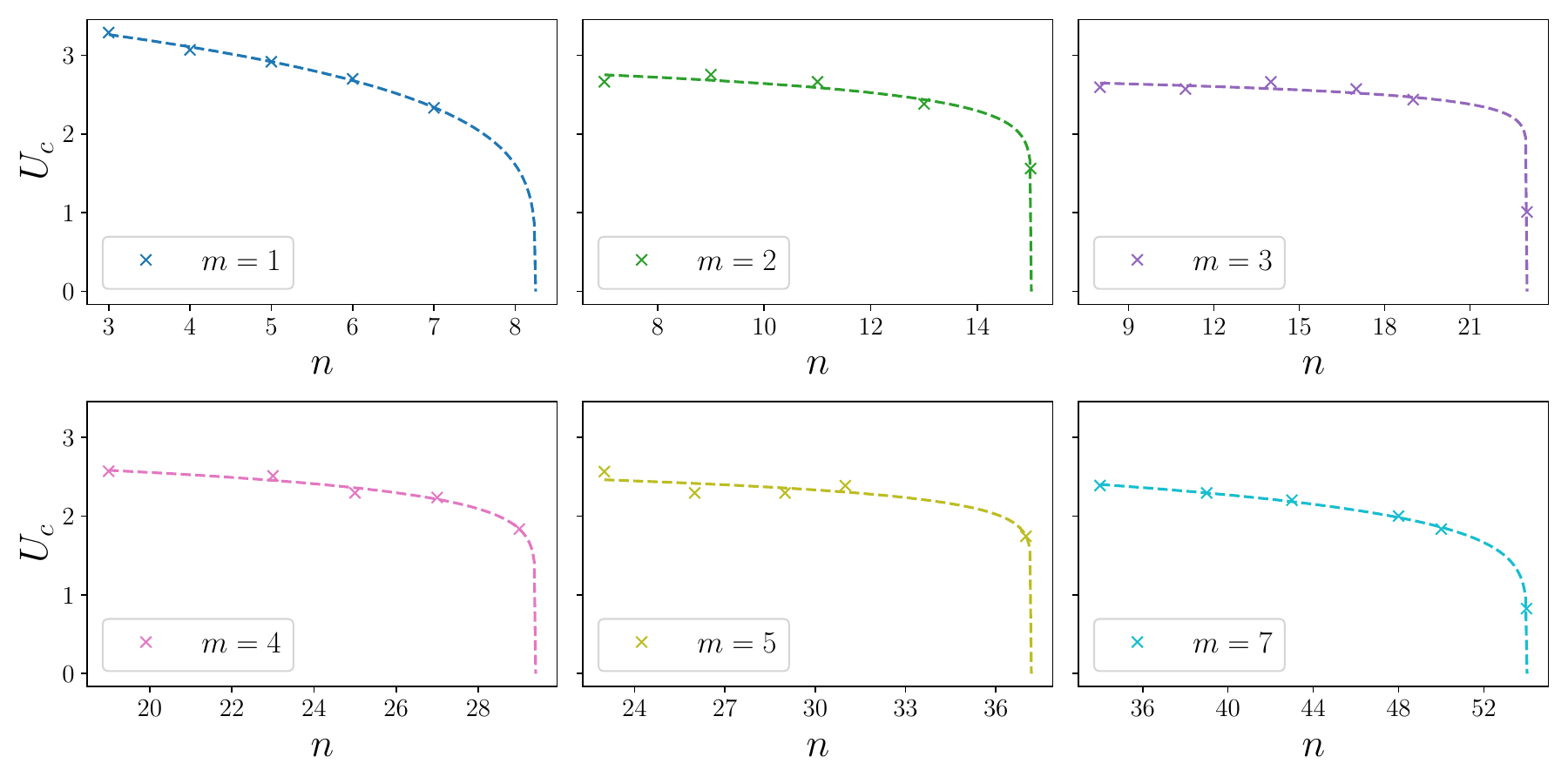}
		\caption{Plots of $U_c$ versus $n$ at fixed $m$, up to $m=7$. The values of $U_c$ are computed using Eq. \eqref{Uc_equation}, with a precision on the integral evaluation up to the fourth digit. The fits (colored dashed lines) are performed according to Eq. \eqref{eq:Uc_various m_fit}.}
		\label{fig:Uc_vs_m_subplots}
	\end{figure}
        
    Even for $m>1$ the scaling of both $\Delta$ and $\tilde{\mu}$ is consistent with the universal exponents $\beta=1/2$ and $\alpha=1$, allowing to conclude that increasing $m$, i.e., changing the number of degenerate bands in the HH model does not modify the nature of the quantum phase transition. However, in practice the scaling regime is more difficult to resolve numerically. The main reason is that the effective cut-off $\Lambda$ defining the region of validity of the parabolic behavior for the DOS around the Weyl point (see Appendix \ref{pairing_field_scaling}) is considerably smaller than in the case $m=1$. As a consequence, the energy window in which the DOS is quadratic is much narrower: since the scaling regime requires $\Delta\ll\Lambda$, reducing $\Lambda$ implies a correspondingly smaller $\Delta$ to resolve the criticality. This makes the required energy resolution extremely demanding. The numerical sampling needed to properly resolve the structure of the DOS around the Weyl node rapidly becomes prohibitive for increasing $m$. Moreover, the solutions of the self-consistent equations appear artificially rounded close to the transition for $\Phi>\Phi_c$, where we expect $U_c\neq0$. This rounding masks the critical behavior, unless very fine grids (both in energy and interaction strengths) are used.

\begin{figure}[t]
		\centering
        \includegraphics[width=0.5\linewidth]{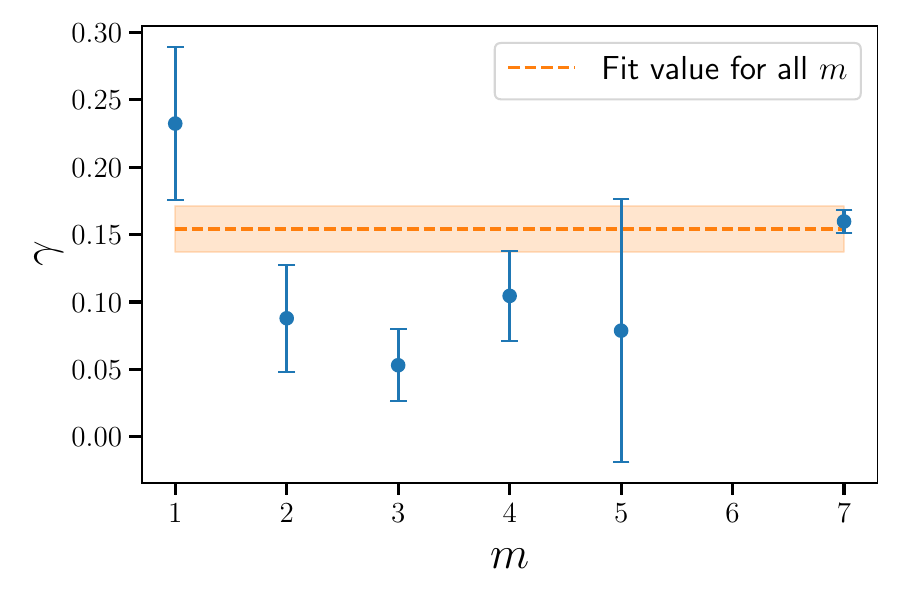}
		\caption{Fit estimates of $\gamma$, using Eq. \eqref{eq:Uc_various m_fit} as a function of $m$. We highlight the value of $\gamma=0.15(2)$ (orange dashed line), estimated in the main text through Eq. \eqref{Uc_vs_phi_gamma_def}, and taking into account all the coprime pairs.}
		\label{fig:gamma_vs_m}
	\end{figure}

    To obtain reliable estimates of the critical interaction strength $U_c$, we exploit the sampled DOS $\rho(\epsilon)$ in Eq. \eqref{Uc_equation}. This procedure is numerically more stable if compared to the vanishing of the pairing field using the self-consistent equations, since $\Delta$ becomes extremely small. The resulting values of $U_c$, at fixed $m$, are then fitted with the expression
    \begin{equation}
        U_c=U_0(n_c-n)^\gamma,
        \label{eq:Uc_various m_fit}
    \end{equation}
    with $U_0$, $n_c$ and $\gamma$ free parameters. We expect $U_0$ and $n_c$ to be $m$-dependent, while $\gamma$ is universal and compatible with the estimate of the main text. The estimates of the fit parameters are reported in Table \ref{table:powerlaw_fit_results}, and plotted for different $m$ in Fig. \ref{fig:Uc_vs_m_subplots}. They turn out  consistent with the previous literature \cite{FontanaPRB2024}. Our results for $\gamma$ confirm that, although the accessible scaling window shrinks progressively with increasing $m$, the critical exponent remain unchanged (see Fig. \ref{fig:gamma_vs_m}).

    Overall, increasing $m$ has therefore a quantitative, rather than qualitative, effect: it reduces the UV cut-off $\Lambda$ and makes the numerical resolution of the scaling regime more demanding, but it does not alter the universal MF exponents.
	
    \subsection{\label{stability_fit_parameters}
    Stability of the fit parameters}
    We further analyze the stability of the fit parameters in Table \ref{table:powerlaw_fit_results}, used to describe the scaling of $U_c$, close to the critical flux $\Phi_c$. The scaling form is expected to be valid asymptotically close to $\Phi_c$. However, performing a reliable fit requires a sufficiently large
    number of data points. It is therefore necessary to verify that including flux values progressively far from the critical point does not significantly affect the estimate of the fitting parameters.

     \begin{figure}[t]
		\centering
        \includegraphics[width=0.8\linewidth]{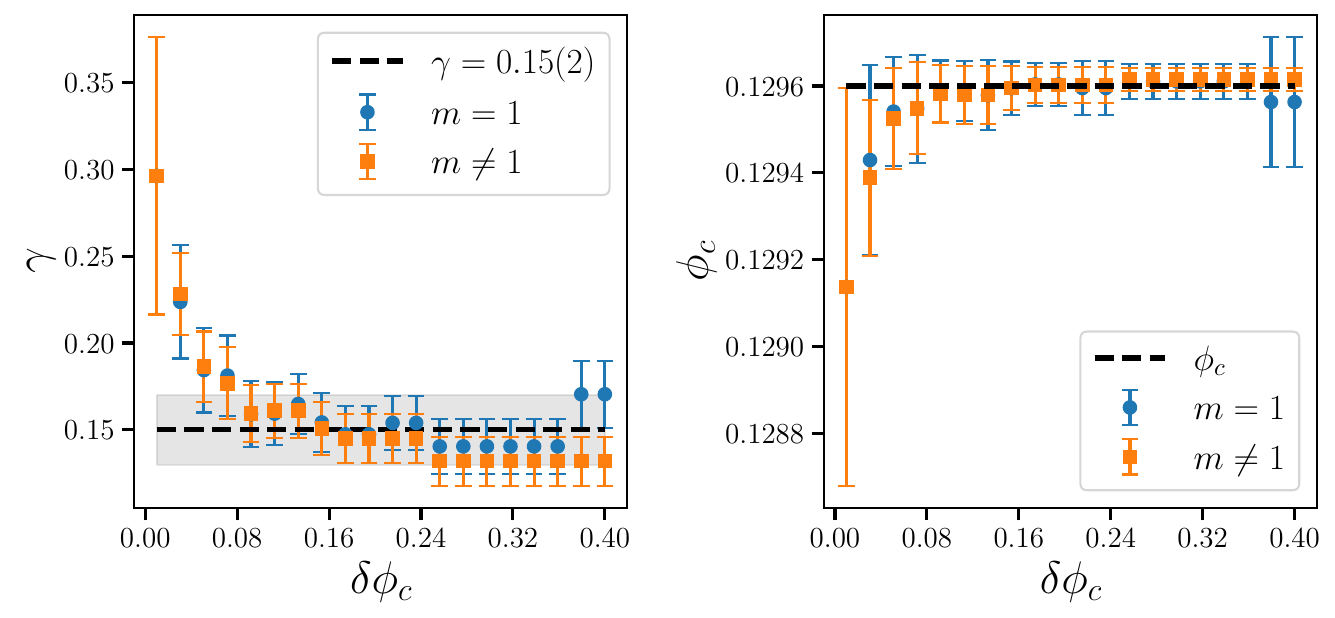}
		\caption{Stability analysis for $\gamma$ (left plot) and $\Phi_c=2\pi n_c/m$ (right plot), as a function of $\delta \Phi_c$. We highlight the corresponding estimates taking into account the data for $m=1$ (blue markers), and all the other data with $m\neq1$ (orange markers). In both the plots, we show the predicted data computed taking into account all the $m$ values (black dashed lines), i.e. the fit parameter $\gamma=0.15(2)$ estimated in the main text, and the critical flux predicted in Ref. \cite{FontanaPRB2024}.}
		\label{fig:gamma_phi_c_stability}
	\end{figure}

    To address this point, we consider different flux windows around $\Phi_c$. Specifically, we define the distance from the critical flux as $\delta \Phi_c\equiv |\Phi-\Phi_c|$, and include in the fit only those flux values below a fixed value of $\delta \Phi_c$. Starting from flux values close to $\Phi_c$, we progressively increase the value of $\delta \Phi_c$, including additional data points. For each choice, we repeat the fit and extract the corresponding values of the fitting parameters. In this way, we monitor the behavior of $\gamma$ and $\Phi_c$ (or equivalently $n_c$), as functions of $\delta \Phi_c$. If the adopted scaling form correctly captures the behavior of the system for $\Phi\simeq\Phi_c$, there must be a window where the fitted parameters remain stable, at least up to a certain distance from $\Phi_c$. Conversely, strong variations of the parameters would signal the possible breakdown of Eq. \eqref{eq:Uc_various m_fit}.

    The results are summarized in Fig. \ref{fig:gamma_phi_c_stability}, where the estimated $\gamma$ and $\Phi_c\equiv2\pi m/n_c$ are plotted, as functions of $\delta \Phi_c$. The emergence of a plateau in a given window indicates that the estimates of the fitting parameters are robust against the inclusion of additional flux values, confirming the stability of the scaling analysis.

    \section{Derivation of the scaling of the pairing field}
    \label{pairing_field_scaling}
    In this Appendix we derive the scaling relation for the pairing field $\Delta$, discussed in Section \ref{scaling_Uc_phi_subsection} of the main text, starting from the zero-temperature gap equation in Eq. \eqref{1st_SCE_T0}. We focus on the case where the DOS vanishes quadratically at the Weyl point,
    \begin{equation}
        \rho(\epsilon)\simeq\rho_0(\epsilon-E_w)^2\qquad\text{for }\epsilon\simeq E_w.
    \end{equation}
    The most significant contribution for the derivation of the scaling comes from the region around the Weyl point. For this reason we focus on the integration range $I_{w}=[E_w-\Lambda,E_w+\Lambda]$, where $\Lambda$ is a high-energy cutoff defining the parabolic approximation, and write Eq. \eqref{1st_SCE_T0} as
    \begin{equation}
        \frac{1}{U}=\frac{\rho_0}{2}\int_{I_w}\;d\epsilon\;\frac{(\epsilon-E_w)^2}{\sqrt{\tilde{\epsilon}^2+\Delta^2}}.
        \label{eq:gap_equation_with_quadratic_DOS}
    \end{equation}
    We further assume that, as a first approximation, the shifted chemical potential $\tilde{\mu}$ equals the Weyl energy, $\tilde{\mu}=E_w$, very close to the phase transition, for $U\simeq U_c$. This is a numerically justified approximation, as we widely discussed in Section \ref{scaling_Uc_phi_subsection}. If the explicit dependence of $\tilde{\mu}$ on $U-U_c$ is considered, i.e. by parametrizing $\tilde{\mu}=E_w-M(U)$, there are subleading corrections to the final scaling relations for it (see Appendix \ref{shifted_mu_scaling}).

    We now change variable to $\xi=\epsilon-E_w$, $I_w\rightarrow [-\Lambda,\Lambda]$. Using parity, 
    \begin{equation}
        \frac{1}{U}=\rho_0\int_0^\Lambda\;d\xi\;\frac{\xi^2}{\sqrt{\xi^2+\Delta^2}}.
        \label{eq:gap_eq_with_mutilde_equal_Ew}
    \end{equation}
    The integral can be computed in a closed form:
    \begin{equation}
        \mathcal{I}(\Delta)\equiv \int_0^\Lambda\;d\xi\;\frac{\xi^2}{\sqrt{\xi^2+\Delta^2}}=\frac{1}{2}\bigg[\Lambda\sqrt{\Lambda^2+\Delta^2}-\Delta^2\log\bigg(\frac{\Lambda+\sqrt{\Lambda^2+\Delta^2}}{|\Delta|}\bigg)\bigg].
    \end{equation}
    The gap equation is $1/U=\rho_0 \, \mathcal{I}(\Delta)$, and is exact for finite UV cutoff. We now expand for small pairing field, i.e. $\Delta\ll\Lambda$, and assume $\Delta>0$ without loss of generality. Using $\sqrt{\Lambda^2+\Delta^2}=\Lambda+\Delta^2/2\Lambda+O(\Delta^4)$ and $\log[(\Lambda+\sqrt{\Lambda^2+\Delta^2})/\Delta]=\log(2\Lambda/\Delta)+O(\Delta^2)$, the gap equation gives
    \begin{equation}
        \frac{1}{U}=\rho_0\frac{\Lambda^2}{2}-\rho_0\frac{\Delta^2}{2}\log\bigg(\frac{2\Lambda}{\Delta}\bigg)+O(\Delta^2).
    \end{equation}
    From Eq. \eqref{Uc_equation} we observe that $1/U_c=\rho_0\Lambda^2/2$, and using this definition and separating the UV-dependent pieces from the $\Delta$-dependent ones, we obtain
    \begin{equation}
        \frac{1}{U}-\frac{1}{U_c}=\rho_0\Delta^2\log\Delta+O(\Delta)^2,
    \end{equation}
    where UV cutoff-dependent terms have been reabsorbed into the definition of $U_c$. To extract the asymptotic dependence of the pairing field on $U-U_c$, we employ a self-consistent iteration where we first ignore the logarithmic factor, and then we replace the obtained estimate into the logarithm and solve for $\Delta$. With this procedure we get
    \begin{equation}
        |\Delta|\sim\sqrt{\frac{U-U_c}{|\log(U-U_c)|}},
        \label{eq:marginal_log_scaling}
    \end{equation}
    capturing the marginal logarithmic correction arising from the quadratic DOS, while the MF exponent is $\beta=1/2$.

    We finally remark that the logarithmic correction arises from the intermediate energy window $\Delta\ll\xi\ll\Lambda$. If one instead expands the integrand asymptotically before performing the exact integration, the logarithmic term is missed, yielding the pure MF behavior $\Delta\propto |U-U_c|^{\frac{1}{2}}$.

    \section{\label{shifted_mu_scaling}Scaling of the shifted chemical potential}
    We show how to derive the scaling relation for the shifted chemical potential $\tilde{\mu}$, discussed numerically in Section \ref{scaling_Uc_phi_subsection} of the main text. We start from the zero-temperature particle number equation in Eq. \eqref{2nd_SCE_T0}, which can be written as
    \begin{equation}
        1-f=\int_{-\epsilon_0}^{\epsilon_0}\;d\epsilon\;\rho(\epsilon)\frac{\tilde{\epsilon}}{\sqrt{\tilde{\epsilon}^2+\Delta^2}} \, .
    \end{equation}
    We focus on the case of $\rho(\epsilon)\simeq\rho_0(\epsilon-E_w)^2$ for $\epsilon\simeq E_w$, since we are interested in the critical behavior close to the Weyl point. This parabolic approximation is valid only for $|\epsilon-E_w|<\Lambda$, where $\Lambda$ is the UV cutoff defining the critical region. By parametrizing $\tilde{\mu}=E_w-M$, and changing integration variable to $\tilde{\epsilon}$, we obtain the $\epsilon-E_w=\tilde{\epsilon}-M$ and the right-hand side of the particle number equation becomes proportional to the integral
    \begin{equation}
        \mathcal{J}_\Lambda(M,\Delta)\equiv\int_{-\Lambda+M}^{\Lambda+M}\;d\tilde{\epsilon}\;\frac{\tilde{\epsilon}(\tilde{\epsilon}-M)^2}{\sqrt{\tilde{\epsilon}^2+\Delta^2}} \, .
        \label{RHS_number_eq}
    \end{equation}
     The particle number equation becomes then $1-f=\rho_0 \, \mathcal{J}_\Lambda(M,\Delta)$. The integral $\mathcal{J}_\Lambda(M,\Delta)$ can be computed exactly for arbitrary $M$. However, we notice that it satisfies the property
     \begin{equation}
         \mathcal{J}_\Lambda(M,\Delta)=\int_{-\Lambda}^{\Lambda}\;d\tilde{\epsilon}\;\frac{\tilde{\epsilon}(\tilde{\epsilon}-M)^2}{\sqrt{\tilde{\epsilon}^2+\Delta^2}}+\frac{2\Lambda^3}{\sqrt{\Lambda^2+\Delta^2}}M+O(M^3)
     \end{equation}
     for $M\ll\Lambda$ (see Section \ref{scaling_Uc_phi_subsection} and Fig. \ref{fig:tilde_mu_vs_U-Uc_Hasegawa_n}).
    Expanding the numerator and keeping only even terms (since the domain of integration is now even), we are left with one integral to be computed, 
     \begin{equation}
        1-f=-2 \, \rho_0 \, M \bigg[\int_{-\Lambda}^{\Lambda}\;d\tilde{\epsilon} \frac{\tilde{\epsilon}^2}{\sqrt{\tilde{\epsilon}^2+\Delta^2}} + {\frac{2\Lambda^3}{\sqrt{\Lambda^2+\Delta^2}}}\bigg] \, .
     \end{equation}
     The result can be written in a closed form, since
     \begin{equation}
        2\int_0^\Lambda\;d\tilde{\epsilon} \frac{\tilde{\epsilon}^2}{\sqrt{\tilde{\epsilon}^2+\Delta^2}}=\Lambda\sqrt{\Lambda^2+\Delta^2}+\frac{\Delta^2}{2}\log\bigg(\frac{\sqrt{\Lambda^2+\Delta^2}-\Lambda}{\sqrt{\Lambda^2+\Delta^2}+\Lambda}\bigg)\, .
     \end{equation}
     We can now further expand for $\Delta\ll\Lambda$, assuming positive pairing fields without any loss of  generality. Since $2\Lambda^3/\sqrt{\Lambda^2+\Delta^2}=(2\Lambda^2-\Delta^2)M+O(\Delta^3)$, we obtain
     \begin{equation}
         1-f=-2 \, \rho_0 \, \bigg[3\Lambda^2+\Delta^2\log\frac{\Delta}{2\Lambda}+O(\Delta)^2\bigg]\, M \,.
     \end{equation}
     Isolating $M$,
     \begin{equation}
         M \simeq M_0\bigg[1-\frac{\Delta^2}{\Lambda^2}\log\frac{\Delta}{2\Lambda}\bigg] \, ,
     \end{equation}
     where $M_0\equiv-(1-f)/ \big(6 \, \rho_0 \, \Lambda^2 \big)$, and in the last step we used that $\Delta^2/\Lambda^2\log(\Delta/2\Lambda)\rightarrow0$ for $\Delta/\Lambda\ll1$.
     
     Using finally the scaling behavior for the pairing field, Eq. \eqref{eq:marginal_log_scaling}, we observe that $\Delta^2\log\Delta\sim(U-U_c)+O\,[(U-U_c)\log(U-U_c)]$, allowing to conclude that
     \begin{equation}
         M\sim M_0 + M_1 \, (U-U_c) \, ,\qquad M_1\equiv-\frac{M_0}{\Lambda^2} \, .
         \label{eq:scaling_M_mu_tilde}
     \end{equation}
     This shows that, within the approximation $M\ll\Delta\ll\Lambda$, valid in the Weyl regime, the MF exponent for the renormalized chemical potential $\tilde{\mu}$ is $\alpha=1$.

     \subsection*{Including corrections to the pairing field scaling}
     
     We briefly discuss here the inclusion the scaling form for the renormalized chemical potential, presented in Eq. \eqref{eq:scaling_M_mu_tilde}, in the derivation of the scaling for the pairing field $\Delta$.

     The starting point is Eq. \eqref{eq:gap_equation_with_quadratic_DOS}. With the change of variables $\xi=\epsilon-E_w$, $\tilde{\mu}=E_w-M$, it reads
     \begin{equation}
         \frac{1}{U}=\frac{\rho_0}  {2} \, \int_{-\Lambda}^{\Lambda}\;d\xi \, \frac{\xi^2}{\sqrt{(\xi+M)^2+\Delta^2}}\equiv\frac{\rho_0}{2}\, \mathcal{K}_\Lambda(M,\Delta).
     \end{equation}
     Contrarily to Eq. \eqref{eq:gap_eq_with_mutilde_equal_Ew}, here we cannot take advantage of parity in $\xi$. Nonetheless, we can expand the integral $\mathcal{K}_\Lambda(M,\Delta)$ for small $M$: the linear term in $M$ is an odd function of $\xi$, leading to a vanishing contribution, since integrated in symmetric domain. We are left with
     \begin{equation}
        \mathcal{K}_\Lambda(M,\Delta)=\mathcal{K}_\Lambda(0,\Delta)+\frac{M^2}{2}\int_{-\Lambda}^\Lambda\;d\xi\;\frac{\xi^2 \, (2\xi^2-\Delta^2)}{(\xi^2+\Delta^2)^{5/2}} \, .
     \end{equation}
     The integral has the closed form
     \begin{equation}
         \int_{-\Lambda}^\Lambda\;d\xi\;\frac{\xi^2(2\xi^2-\Delta^2)}{2(\xi^2+\Delta^2)^{5/2}}=\log\frac{\sqrt{\Lambda^2+\Delta^2}+\Lambda}{\sqrt{\Lambda^2+\Delta^2}-\Lambda}-\frac{3\Lambda^3+2\Lambda\Delta}{(\Lambda^2+\Delta^2)^{3/2}}\simeq\log\frac{\Lambda}{\Delta}+O(\Delta^2),
     \end{equation}
     up to $\Lambda$-dependent terms constant in $\Delta$. This means that $\mathcal{K}_\Lambda(M,\Delta)=\mathcal{K}_\Lambda(0,\Delta)+M^2[\log(\Lambda/\Delta)+O(\Delta^2)]$.

     Inserting back in the gap equation, when we consider the contribution $\mathcal{K}_\Lambda(0,\Delta)$, we recover the same steps done in Appendix \ref{pairing_field_scaling}, with an additional term quadratic in $M$:
     \begin{equation}
         \frac{1}{U}-\frac{1}{U_c}=\rho_0\bigg[\Delta^2\log\Delta+\frac{M^2}{2}\log\frac{\Lambda}{\Delta}\bigg].
     \end{equation}
     Since we just showed that $M\sim U-U_c$ at the leading order, we conclude that the additional term is $\propto (U-U_c)^2$. Close to the quantum phase transition, where $U\rightarrow U_c$, the first term in the right-hand side dominates \footnote{This is true unless particular cases for which the logarithmic term compensate the quadratic dependence in $U-U_c$, depending on the interplay of the UV cutoff and the pairing field in the parabolic region, a behavior that we do not observe for the cases considered in this paper.}, supporting our numerical evidence of Section \ref{results_section}.
    
\end{document}